\documentclass[10pt,journal,letterpaper,compsoc]{IEEEtran}

\usepackage{amssymb}

\ifCLASSINFOpdf

\else

  \usepackage[dvips]{graphicx}

  \DeclareGraphicsExtensions{.eps}
\fi

\usepackage{algorithm}
\usepackage{algorithmic}

\begin{document}

\title{Topological Centrality and Its Applications}

\author{Hai~Zhuge,~\IEEEmembership{Senior Member,~IEEE,}
  and~Junsheng~Zhang%
  \IEEEcompsocitemizethanks{\IEEEcompsocthanksitem The authors are with China
    Knowledge Grid Research Group, Key Lab of Intelligent Information
    Processing, Institute of Computing Technology, and graduate school, Chinese
    Academy of Sciences, 100190, PO Box 2704-28, Beijing, China.
    \protect\\
  
  }%
  \thanks{}}

\IEEEcompsoctitleabstractindextext{%
  \begin{abstract}
    Recent development of network structure analysis shows that it plays an
    important role in characterizing complex system of many branches of
    sciences. Different from previous network centrality measures, this paper
    proposes the notion of topological centrality (TC) reflecting the
    topological positions of nodes and edges in general networks, and proposes
    an approach to calculating the topological centrality. The proposed
    topological centrality is then used to discover communities and build the
    backbone network. Experiments and applications on research network show the
    significance of the proposed approach.
  \end{abstract}

  \begin{IEEEkeywords}
    Network structure, Centrality, Community, e-Science
  \end{IEEEkeywords}}

\maketitle

\IEEEdisplaynotcompsoctitleabstractindextext

\IEEEpeerreviewmaketitle

\section{Introduction}

\IEEEPARstart{T}{he} rich get richer phenomenon exists in many complex networks
like the World Wide Web. It is known that there are two ways for a node to
become richer: connecting to more nodes; and, connecting to more important
nodes.

We observe that \textit{a node may earn more if it connects to an important node
  than connects to many but less important nodes, and that both nodes and edges
  play an important role in forming network centrality.}

Existing centrality measures focus on nodes. They cannot explain the topological
characteristic of centrality. This paper is to explore a new network centrality
called topological centrality.

Various centrality measures are defined in a graph $G= (V, E)$, where $V$ is the
vertex set, $E$ is the edge set, $|V|=n$, and $|E|=m$.

The \textit{authority} and \textit{hub} reflect in-degree and out-degree
characteristics of a node in the Web respectively
\cite{DBLP:journals/jacm/Kleinberg99}. The idea of HITS is that a good hub links
to many authorities, while a good \textit{authority} is linked by many good
\textit{hubs}. Nodes with the highest authority or hub in the Web graph act as
authority centers and hub centers. The \textit{authority} and \textit{hub} of a
node are calculated by:
\[
\left\{ \begin{array}{l}
    a(i) = \sum\limits_{(j,i) \in E} {h(j)}  \\
    h(j) = \sum\limits_{(i,j) \in E} {a(i)}  \\
  \end{array} \right. ,
\]
where $a(x)$ and $h(x)$ are the authority and hub of node $x \in \{i, j \}$
respectively.

\textit{Degree centrality} describes the degree information of each node
\cite{freeman1979csn} \cite{nieminen1974cg}. It is based on the idea that more
important nodes are more active, that is, they have more neighbors in the graph.
Degree centrality can be used to find the core nodes of a community; however, it
only considers the \textit{hub} characteristic and ignores the
\textit{authority} characteristic. Degree Centrality $C_{D}(v)$ for a vertex $v$
is calculated as follows:
\[
C_{D}(v)=\frac{deg(v)}{n-1}.
\]

Calculating degree centrality for all nodes $V$ in a graph takes $O(n^{2})$ in a
dense adjacency matrix representation of the graph. While in a sparse graph with
edges $E$, the time complexity is $O(m)$. Similar to the degree centrality, an
approach was proposed to improve the efficiency of information propagation in
P2P network based on the in- and out-degrees of nodes \cite{zhuge2007ppm}.

\textit{Betweenness centrality} describes the frequencies of nodes in the
shortest paths between two indirectly connected nodes \cite{freeman1979csn}
\cite{anthonisse1971rg} \cite{freeman1977smc}. It is based on the idea that if
more nodes are connected via a node, then the node is more important.
Betweenness centrality can be used to find the edges between two communities in
a complex network. Betweenness Centrality $C_{B}(v)$ for vertex $v$ is:
\[
C_B (v) = \sum\limits_{ s \ne v \ne t \in V \atop s \ne t } {\frac{{\sigma _{st}
      (v)/\sigma _{st} }}{{(n - 1)(n - 2)}}},
\]
where $\sigma_{st}$ is the number of shortest geodesic paths from $s$ to $t$,
and $\sigma_{st}(v)$ is the number of shortest geodesic paths from $s$ to $t$
that pass through a vertex $v$. The shortest paths between each pair of nodes in
a graph can be found by Floyd-Warshall algorithm with time complexity $O(n^{3})$
\cite{321107}, so the time complexity of betweenness centrality is also
$O(n^{3})$. Betweenness centrality has been used to study community structure of
social and biological networks \cite{girvan2002css}.

\textit{Closeness centrality} describes the efficiency of the information
propagation from one node to the other nodes \cite{freeman1979csn}
\cite{sabidussi1966cig} \cite{wasserman1994sna}. It is based on the idea that if
a node can quickly reach others, then the node is central. Closeness centrality
can be regarded as a measure of how long it will take information to spread from
a given vertex to other reachable vertices in the network. Closeness Centrality
is defined as the mean geodesic distance (i.e., the \textit{shortest path})
between a vertex $v$ and all other vertices reachable from $v$:
\[
C_c (v) = \frac{{n - 1}}{{\sum\limits_{t \in V\setminus v} {d_G (v,t)} }},
\]
where $n \geq 2$ is the size of the network's connected component reachable from
$v$. Calculating the closeness centrality for each node in the graph has time
complexity $O(n^{3})$.

\textit{Eigenvector centrality} describes the importance of nodes according to
the adjacent matrix of a connected graph \cite{bonacich1972faw}. It assigns
relative scores to all nodes in the network based on the principle that
connections to high-scored nodes contribute more to the score of a node than
connections to low-scored nodes. \textit{PageRank} is a variant of the
eigenvector centrality measure \cite{larry1998pcr}.

\textit{Information centrality} describes nodes' influence on the
\textit{network efficiency} of information propagation \cite{latora2004mcb}. The
network efficiency is defined by
\[
E_{G}=\frac{\sum_{i \ne j \in G}\epsilon_{ij}}{n(n-1))} = \frac{1}{n(n-1)}\sum
\limits_{i \ne j \in G}{\frac{1}{d_{ij}}},
\]
where the efficiency $\epsilon_{ij}$ in the communication between two points $i$
and $j$ is equal to the inverse of the shortest path length $d_{ij}$. The
information centrality of a vertex $i$ is defined as the relative drop in the
network efficiency caused by the removal from $G$ of the edges incident with
$v$:
\[
C_{I}(v)=\frac{\Delta E}{E} = \frac{E[G]-E[G_{v}']}{E},
\]
where $G_{v}'$ indicates a network by removing the edges incident with node $v$
from $G$. Information centrality has been used to study the structures of
communities in complex networks \cite{fortunato2004mfc}.

\section{Topological Centrality}

\subsection{Definition}

In a dynamic network, the weights of nodes and the weights of edges will
influence each other and keep changing. Each time of influence between each pair
of nodes is called one time of \textit{iteration}. If the order of nodes'
weights keeps unchanging after many times of iteration, the network reaches the
\textit{steady state} and the nodes with the highest weights are called
\textit{topological centers}. An undirected graph may have one or more
topological centers. The number of topological centers is decided by the graph
structure. An undirected network may have one of the following structures.

\begin{figure}[!htb]
  \centering
  \resizebox{0.95\columnwidth}{!}{\includegraphics{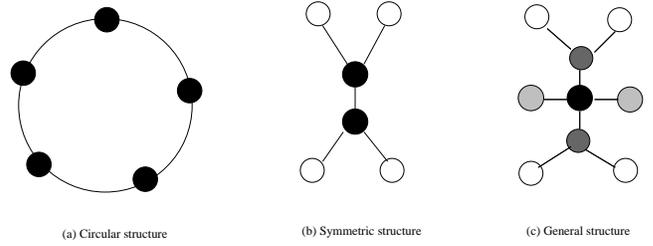}}
  \caption{Three types of topological structures. The darker is the node, the
    higher the topological centrality is. The black nodes are the topological
    centers. Networks of circular structure have $n~(n \geq 3)$ topological
    centers; network of symmetric structure has $2$ topological centers; other
    networks have $1$ topological center.}
  \label{fig:topological-centers}
\end{figure}

\begin{enumerate}%
\item A network with \textit{circular} structure has $n~(n \ge 3)$ topological
  centers as shown in Fig. \ref{fig:topological-centers}a.

\item A network with \textit{symmetric} structure has two topological centers as
  shown in Fig. \ref{fig:topological-centers}b.

\item Otherwise, the network has a unique topological center as shown in
  Fig.\ref{fig:topological-centers}c.

\end{enumerate}

In an undirected graph, the length of the shortest path between two nodes in a
graph is the geodesic distance between them. Especially, if two nodes are
unreachable, then their geodesic distance is $+\infty$. Geodesic distance can be
used to find the nearest topological center of a node.

\textit{When a network is in the steady state, the topological centrality (TC)
  of a node is the ratio of its weight to the largest weight of nodes. The
  topological centers have the largest weight of node $1$. The topological
  centrality of an edge is the ratio of its weight to the largest weight of
  node.}

The TC of a node reflects the geodesic distance from a node to its nearest
topological center. The TC of an edge reflects the geodesic distance from the
edge to its nearest topological center. The higher is the TC of a node/edge, the
closer it is to the nearest topological center.

\subsection{Calculating Topological Centrality}

\textbf{Hypothesis 1.} \textit{The topological centrality of a node is
  positively influenced by the topological centrality degrees of its neighbor
  nodes.}

Hypothesis $1$ leads to the following characteristics:
\begin{enumerate}%
\item a node connecting to nodes with higher TC degrees gets higher TC degree;
  and,
\item a node connecting to more nodes gets higher TC degree.
\end{enumerate}

\textbf{Hypothesis 2.} \textit{If two nodes of an edge have higher TC degrees,
  then the edge has higher TC; and, if an edge has higher TC, then its two nodes
  also have higher TC degrees.}

Hypothesis $2$ leads to the following characteristics:
\begin{enumerate}%
\item nodes closer to the topological center have higher TC degrees; and,

\item edges closer to the topological center have higher TC degrees. These
  characteristics reflect that nodes with higher TC degrees are incident with
  edges having higher TC degrees.
\end{enumerate}

The two hypotheses can be represented by:
\begin{equation}
  \left\{ \begin{array}{l}
      \omega (n) \uparrow  = \omega (n) + \sum {g(\omega (link(n,n_i )) \uparrow ,\omega (n_i ) \uparrow )}  \\ 
      \omega (l) \uparrow  = f(\omega (l_s ) \uparrow ,\omega (l_t ) \uparrow ) \\ 
    \end{array} \right. 
\end{equation}

where $n$ is a node, $n_{i}$ are neighbors of $n$, $\omega(link(n,n_{i}))$ is
the weight of link between $n$ and $n_{i}$; $l$ is a link, $l_{s}$ and $l_{t}$
are the source and target nodes of $l$ respectively; $f$ and $g$ are two
functions, and $\uparrow$ means the positive correlative relations.

During the calculation process of TC degree, the weights of nodes and edges will
increase after each time of iteration, but the descending order of weights of
nodes will converge to the steady state. The weights of nodes can be normalized
by dividing the largest weight of nodes. If the normalized weights of nodes
converge, the descending order of nodes' weights will keep unchanging, and the
edges' weights will also converge. The converged nodes' weights and edges'
weights are the TC degrees of nodes and links respectively.

Normalization of weights of nodes satisfies the following characteristics:
\begin{enumerate}%
\item If the normalized weights of nodes converge, then the order of nodes by
  descending the weights of nodes will also converge. The normalization process does
  not change the order of weights of nodes. The difference is that the weights of nodes
  are mapped onto the interval $(0, 1]$.
\item If the normalized weights of nodes converge, the weights of edges also
  converge. According to the definition of TC of an edge, the weights of edges are
  the sum of the weights of its two end nodes. Since the normalized weights of
  nodes converge, the weights of incident edges will also converge.
\item If the normalized weights of nodes converge, then the TC degrees of edges
  converge. It is also obvious, because the normalization of weights of edges is
  just to map the weights of edges onto the interval $(0, 1]$, and keeps the
  order of weights of edges.
\end{enumerate}

We propose the following approach to calculating the TC in a connected network.
Suppose a connected graph $G = (V, E)$ with $n~(n > 1)$ nodes and $m~(m \geq n -
1)$ edges, $V = {v_{1},v_{2},\ldots, v_{n}}$, $E = {e_{1},e_{2},\ldots,e_{m}}$,
and the corresponding adjacency matrix is $A$. The element of $A$ is $a_{ij}$,
and,

\[
a_{ij} = \left\{ \begin{array}{ll}
    1 & \{ i,j\}  \in E \\
    0 & \{ i,j\}  \notin E \\
  \end{array} \right. .
\]

The following formula implements the iterative calculation of topological
centrality of nodes and edges, where $temp\_{\omega_{i}}$ and $\omega_{i}$ are
the weights of $v_{i}$ before and after normalization, and
$temp\_{\omega_{e(i,j)}}$ and $\omega_{e(i,j)}$ are the weights of edge $e(i,j)$
before and after normalization, and $t \geq 0$ is the iteration time.

\begin{equation}%
  \left\{
    \begin{array}{ll}
      temp\_{\omega_{i}^{(t+1)}} & = \omega_{i}^{(t)} + \sum_{j=1}^{n}a_{ij}\omega_{e(i,j)}^{(t))}\omega_{j}^{(t)} \\
      temp\_{\omega_{e(i,j)}^{(t+1)}} & = temp\_{\omega_{i}^{(t+1)}} + temp\_{\omega_{j}^{(t+1)}}
    \end{array}
  \right.
\end{equation}

The following formulas normalize the TC degrees of nodes and links.
\begin{equation}%
  \left\{
    \begin{array}{ll}%
      \omega_{i}^{(t+1)} & = \frac{temp\_{\omega_{i}^{(t+1)}}}{Max_{i=1}^{n}temp\_{\omega_{i}^{(t+1)}}} \\
      \omega_{e(i,j)}^{(t+1)} & = \frac{temp\_{\omega_{e(i,j)}^{(t+1)}}}{Max_{j=1}^{m}temp\_{\omega_{e(i,j)^{(t+1)}}}}
    \end{array}
  \right.
\end{equation}

The iterative calculation terminates, if the following conditions are satisfied:
\begin{equation}%
  \left\{
    \begin{array}{l}
      \sum_{i=1}^{n}(\omega_{i}^{(t+1)}-\omega_{i}^{(t)})^{2} < \epsilon_{N} \\
      \sum_{j=1}^{m}(\omega_{e_{j}}^{(t+1)}-\omega_{e_{j}}^{(t)})^{2} < \epsilon_{M}
    \end{array}
  \right.
\end{equation}

Algorithm \ref{alg:tc-calculation} calculates the weights of nodes and links
iteratively, where $MAX$, $\epsilon_{N}$ and $\epsilon_{M}$ control the times of
iterative calculation.

\begin{algorithm*}[!htb]%
  \centering
  \caption{Calculating topological centrality degrees of nodes and edges}
  \label{alg:tc-calculation}
  \begin{algorithmic}[1]%
    \REQUIRE node number $n$, edge number $m$, edges like $(linknum, starNode,
    endNode, weight)$, limited iteration time $MAX$, deviation square limit of
    weight difference of nodes $\epsilon_{N}$, deviation square limit of weight
    difference of links $\epsilon_{M}$;

    \STATE $nodeWeight[1..n] \leftarrow 1, count \leftarrow 0, nodeSum
    \leftarrow n, edgeSum \leftarrow m$

    \WHILE{ ($count < MAX$) and (( $nodeSum > \epsilon_{N}$ ) or ($edgeSum >
      \epsilon_{M}$)) } \STATE $oldNodeWeight[1..n] \leftarrow nodeWeight[1..n]$
    \STATE $oldEdgeWeight[1..m] \leftarrow edgeWeight[1..m]$ \STATE
    $nodeWeight[1..n] \leftarrow
    \frac{nodeWeight[1..n]+\sum_{incident~edge}edgeWeight*nodeWeight}{max(nodeWeight)}
    $ \STATE $edgeWeight \leftarrow \frac{\sum
      _{inciden~node}nodeWeight}{max(edgeWeight)}$ \STATE $nodeSum \leftarrow
    \sum_{i= 1}^{n}(nodeWeight[i]-oldNodeWeight[i])^{2}$ \STATE $edgeSum
    \leftarrow \sum_{i=1}^{m}(edgeWeight[i]-oldEdgeWeight[i])^{2}$ \STATE $count
    \leftarrow count + 1$
    \ENDWHILE
    \RETURN $nodeWeight[1..n]$ and $edgeWeight[1..m]$
  \end{algorithmic}
\end{algorithm*}

The time complexity of Algorithm \ref{alg:tc-calculation} is $O(MAX (n + m))$.
At the initializing stage, all the weights of nodes are assigned to $1$. If the
weights of edges are not given, then all the weights of edges are assigned $1$.
After the first iteration, the weight of a node in next iteration is the sum of
weights of its neighbor nodes and its own weight; then the weights of edges are
the sum of two end nodes. The values of weights of nodes become larger comparing
to the initial values. The weights of nodes and edges are normalized by dividing
the maximum weight of nodes and edges during each time of iteration.

Algorithm \ref{alg:tc-calculation} has two termination conditions: one is the
maximum iteration times $MAX$; the other is the square deviation threshold of
weight difference of nodes $\epsilon_{N}$ and the square deviation threshold of
weight difference of edges $\epsilon_{M}$. After Algorithm
\ref{alg:tc-calculation} stops, the nodes with weights \ref{alg:tc-calculation}
are the topological centers. The weight of a node is topology centrality, and
the larger is the weight of node, the closer the node is to the nearest
topological center.

Table \ref{tab:centrality-comparison} makes a comparison between the topological
centrality and other centrality measures.

\begin{table}[!htb]
  \centering
  \caption{Comparison of different centrality measures}
  \label{tab:centrality-comparison}
  \begin{tabular}{|l|l|l|}
    \hline
    Centrality Measure & Time Complexity & About Node or Edge \\
    \hline
    \hline
    degree centrality & $O(n^{2})$ & node \\
    \hline
    betweenness centrality & $O(n^{3})$ & node or edge\\
    \hline
    closeness centrality & $O(n^{3})$ & node\\
    \hline
    eigenvector centrality & - & node\\
    \hline
    information centrality  & $O(n^{3})$ & node\\
    \hline
    topological centrality & $O(K(n + m))$ & node and edge \\
    \hline
  \end{tabular}
\end{table}

\subsection{Experiments}

\subsubsection{Convergence Experiment}

\begin{figure*}[!htbp]
  \centering
  \resizebox{\textwidth}{!}{\includegraphics{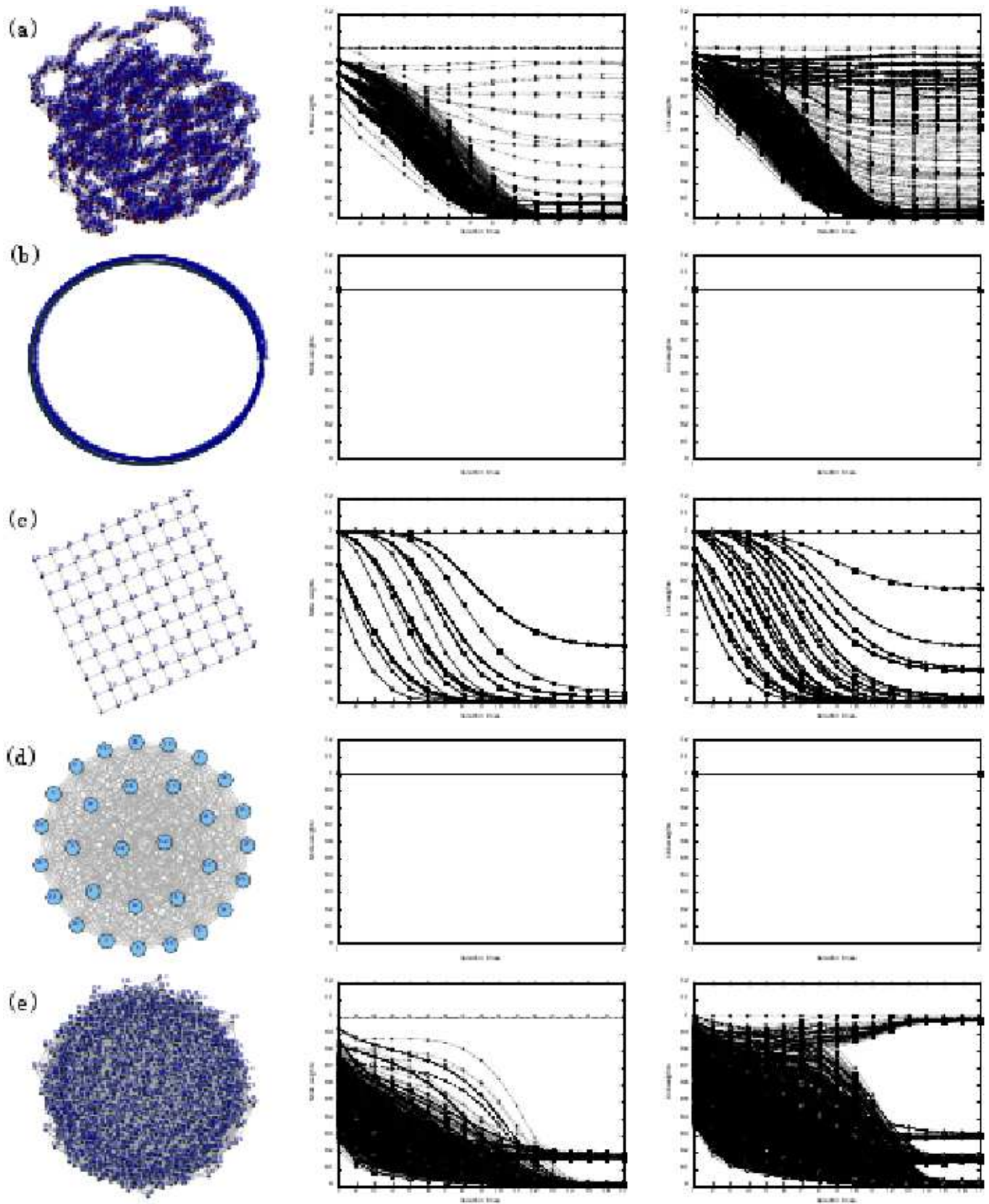}}
  \caption{Topological centrality convergence experiments ($MAX=100$,
    $\epsilon_{N} = \epsilon_{M} = 0.001$): the left column lists networks of
    several structures; the middle column lists the node convergence records
    (x-axis is iteration times, and y-axis is normalized weights of nodes);
    and, the right column lists the link convergence records (x-axis is
    iteration times, and y-axis is normalized weights of links). (a)
    Watts-Strogatz small-world network with $n = 1000$ and $m = 5000$, and
    iteration time is $14$; (b) ring network with $n = 1000$ and $m = 1000$, and
    iteration time is $2$; (c) lattice network with $n = 100$ and $m = 180$, and
    iteration time is $17$; (d) full network with $n = 30$ and $m = 435$, and
    iteration time is $2$; (e) Ed{\"o}rs-R{\'e}nyi random graph with $n =
    1000$, $p = 0.02$, $m = 10045$, and iteration time is $17$.}
  \label{fig:convergence-exp}
\end{figure*}

We carry out experiments on several types of network to verify the convergence
of the algorithm. Fig. \ref{fig:convergence-exp} shows the experiment results of
iterative TC calculation for nodes and links in different structured networks
with different scales: (a) Watts-Strogatz small-world network with $n = 1000$
and $m = 5000$; (b) ring network with $n = 1000$ and $m = 1000$; (c) lattice
network with $n = 100$ and $m = 180$; (d) full network with $n = 30$ and $m =
435$; (d) Ed{\"o}rs-R{\'e}nyi random graph with $n = 1000$, $p = 0.02$, and $m =
10045$. Experiment results show that the TC degrees of node and links can
converge after many times of iteration, which is related to $n$, $m$,
$\epsilon_{N}$ and $\epsilon_{M}$.

\subsubsection{Comparison of Centrality Measures}

\begin{figure}[!htb]
  \centering
  \resizebox{0.9\columnwidth}{!}{\includegraphics{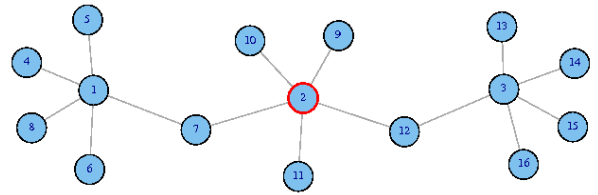}}
  \caption{A simple case (a tree with 16 nodes) for the comparison of centrality
    measures.}
  \label{fig:centrality-compare-case}
\end{figure}

Different centrality measures such as degree centrality, betweenness centrality,
closeness centrality and information centrality are compared in
\cite{zhuge06:_discov_knowl_flow_scien}. Here we add two extra centrality
measures: one is the PageRank of node as an instance of eigenvector centrality,
the other is the topological centrality we proposed. The comparison is based on
Fig. \ref{fig:centrality-compare-case} which is a tree with $16$ vertices. Table
\ref{tab:centrality-compare} shows different centrality degrees of vertices in
Fig. \ref{fig:centrality-compare-case}. The experiment results show the
following characteristics:
\begin{enumerate}%
\item Degree centrality is a local centrality, and it only records the degrees
  of nodes without any global information. Nodes 1, 2, and 3 have degree 5,
  nodes 7 and 12 have the degree 2, and the other nodes have degree 1. Degree
  centrality is normalized by the number of edges 15.
\item Closeness centrality has similar result as information centrality. The
  difference is that the orders of nodes \{1, 3\} and \{7, 12\} are different.
  Information centrality degrees of vertex 1 and 3 are larger than 7 and 12.
  Because information centrality concentrates on the network efficiency. The
  influence on network efficiency by removing 1 and 3 is larger than that by
  removing 7 and 12.
\item PageRank result is far from other measures. Nodes 1 and 3 are two centers
  in PageRank, and node 2 have lower PageRank than nodes 1 and 3, because the
  authority of nodes 7 and 12 are divided into two parts, while nodes 1 and 3
  have four neighbors which contributes all of their authority values to nodes 1
  and 3 respectively. Nodes 7 and 12 have higher rank values than nodes 9, 10
  and 11, because they have more neighbors.
\item Betweenness centrality reflects the frequencies of nodes occurring in the
  shortest paths between indirectly connected node pairs. However, betweenness
  centrality has the worst resolution of nodes. Node 2 has the highest
  betweenness centrality, nodes 1, 3, 7, and 12 have higher betweenness
  centrality, and the others have the same betweenness centrality 0.

\item Topological centrality combines the degree information and neighbor
  weights information. It has the characteristics of degree centrality and
  PageRank. Node 2 is the topological center of the graph. Nodes 7 and 12 have
  higher TC degrees than nodes 9, 10 and 11 because they have extra neighbors.
  Nodes 1 and 3 follow nodes 9, 10 and 11, and then the left vertices. The order
  of node TC degrees confirms the geodesic distance between nodes and the
  topological centers correctly.

\end{enumerate}

\begin{table}[!htb]
  \centering
  \caption{Comparison between topological centrality and other centrality measures}
  \label{tab:centrality-compare}
  \begin{tabular}{|l|l|l|l|l|l|l|}
    \hline
    $v$ & $C_{I}(v)$ & $C_{D}(v)$ & $C_{C}(v)$ & $C_{B}(v)$ & $PR(v)$ & $log(C_{T}(v))$ \\
    \hline
    \hline
    2 & 0.591 & 0.333 & 0.455 & 0.714 & 0.153 & 0.0  \\ \hline
    7 & 0.389 & 0.133 & 0.405 & 0.476 & 0.063 & -0.755\\ \hline
    12 & 0.389 & 0.133 & 0.405 & 0.476 & 0.063 & -0.755\\ \hline
    9 & 0.116 & 0.067 & 0.319 & 0.000 & 0.035 & -0.827\\ \hline
    10 & 0.116 & 0.067 & 0.319 & 0.000 & 0.035 & -0.827\\ \hline
    11 & 0.116 & 0.067 & 0.319 & 0.000 & 0.035 & -0.827\\ \hline
    1 & 0.444 & 0.333 & 0.349 & 0.476 & 0.161 & -2.454\\ \hline
    3 & 0.444 & 0.333 & 0.349 & 0.476 & 0.161 & -2.454\\ \hline
    4 & 0.106 & 0.067 & 0.263 & 0.000 & 0.037 & -5.718\\ \hline
    5 & 0.106 & 0.067 & 0.263 & 0.000 & 0.037 & -5.718\\ \hline
    6 & 0.106 & 0.067 & 0.263 & 0.000 & 0.037 & -5.718\\ \hline
    8 & 0.106 & 0.067 & 0.263 & 0.000 & 0.037 & -5.718\\ \hline
    13 & 0.106 & 0.067 & 0.263 & 0.000 & 0.037 & -5.718\\ \hline
    14 & 0.106 & 0.067 & 0.263 & 0.000 & 0.037 & -5.718\\ \hline
    15 & 0.106 & 0.067 & 0.263 & 0.000 & 0.037 & -5.718\\ \hline
    16 & 0.106 & 0.067 & 0.263 & 0.000 & 0.037 & -5.718 \\ \hline
  \end{tabular}
\end{table}

\subsubsection{Topological Centrality Distributions on Research Network}

Here DBLP dataset is used to study the structure and discover communities in
heterogeneous networks. It contains part of metadata of papers provided by DBLP
in XML formats. The number of papers is $664,188$, and the number of
\textit{citation} relations is $79,128$. The heterogeneous research network is
based on the DBLP data set. The resource types are papers, researchers and
conferences. The semantic links are \textit{authorOf} between researcher and
paper, \textit{coauthor} between researchers, \textit{publishedIn} between paper
and conference/journal, and \textit{cite} between papers.

\begin{figure}[!htb]
  \centering
  \resizebox{0.9\columnwidth}{!}{\includegraphics{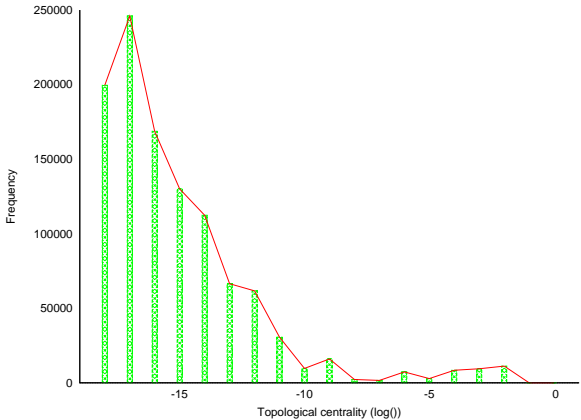}}
  \caption{Topological centrality distributions.}
  \label{fig:tc-distribution}
\end{figure}

The research network contains $1,084,198$ semantic nodes and $2,153,385$
semantic links. The iteration time limits are $MAX = 40$ and $\epsilon_{M} =
\epsilon_{N} = 200$. The distribution of TC degrees of nodes is shown in Fig.
\ref{fig:tc-distribution}. It shows that nodes with lower TC degree contain more
resources than those with higher TC degree.

\section{Application: Discovering Research Communities}

\subsection{Research Community}

Research communities are formed by relations among researchers, papers,
projects, and research activities. Differences between research communities and
graph-based communities are as follows.

\begin{enumerate}%
\item Research communities are dynamically formed by research activities such as
  applying (e.g., funding and position), cooperating, publishing, and citing.
  Communities in general complex networks are viewed from connections (nodes
  within a community are linked more densely than nodes cross communities).

\item Research communities contain multiple types of nodes (researchers and
  papers can play different roles in research activities as discussed in
  \cite{zhuge06:_discov_knowl_flow_scien}) and relations (e.g., coauthor
  relation and citation relation). There are no differences of nodes and edges
  in graph-based communities.
\end{enumerate}

Among existing centrality measures, only the PageRank considers the influences
between neighbor nodes, and the authority of a node is divided by its neighbors.
However, PageRank does not reflect different influences of edges, that is, all
the weights of edges are $1$. In research network: \textit{collaborations
  between authority researchers are more important, and citations between
  authority papers are more important.}

Topological centrality can well distinguish roles of different nodes in research
network. (1) Nodes in a network elect the core nodes by a voting-like mechanism:
\textit{a node connecting to more nodes is more probable to be the local core
  nodes.} After a certain times of iterations, the local core nodes and the
global topological centers are elected. The topological centers are the nodes
connecting to the most core nodes with higher TC degrees. (2) Edges may play
different roles on the mutual influence between the TC degrees of nodes. This
confirms the phenomena of research communities: a researcher cooperating with
authority researchers will be closer to the centers of a research community; a
paper citing (citing may not be true) or is cited by authority papers will be
more possible to be closer to the core papers on a research topic.

\subsection{Roles of Nodes}

Nodes can play different roles according to topological positions in
communities: \textit{core node}, \textit{margin node}, \textit{bridge node} and
\textit{mediated node}.

\begin{enumerate}
\item \textit{Core nodes} are usually hub or authority in the community;
\item Margin nodes belong to one community, and they have few connections to
  other nodes in the community;
\item \textit{Bridge nodes} connect to two or more communities, and they usually
  have equal number of connections to two or more communities; and,
\item Other nodes except the \textit{core nodes}, \textit{margin nodes} and
  \textit{bridge nodes} are \textit{mediated nodes}.
\end{enumerate}

\begin{figure}[!htb]
  \centering
  \resizebox{0.95\columnwidth}{!}{\includegraphics{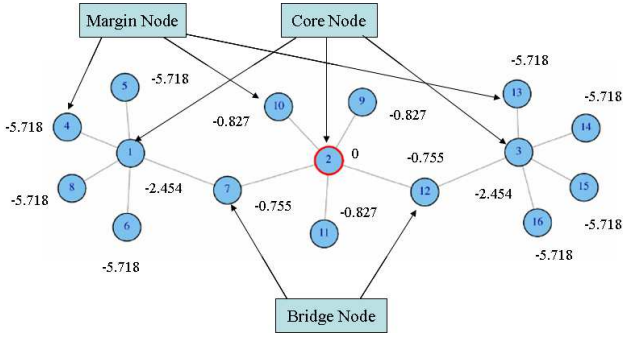}}
  \caption{Distinguishing roles of nodes with topological centrality degrees.}
  \label{fig:node-roles}
\end{figure}

The proposed topological centrality can be used to distinguish roles of nodes.
For example, Fig. \ref{fig:node-roles} contains three communities: $C_1=\{1, 4,
5, 6, 7, 8\}$, $C_2=\{2, 7, 9, 11, 12\}$ and $C_3=\{3, 12, 13, 14, 15, 16\}$.
Node $1$, $2$ and $3$ are the core nodes of $C_1$, $C_2$ and $C_3$ respectively;
Nodes $7$ and $12$ are bridge nodes; nodes $4$, $5$, $6$ and $8$ are margin
nodes of $C_1$; nodes $9$, $10$ and $11$ are margin nodes of $C_2$; and, nodes
$13$, $14$, $15$ and $16$ are margin nodes of $C_3$.

Nodes can be classified by TC degrees.
\begin{enumerate}%
\item If the TC degree of a node is larger than that of most of its neighbors,
  then the node is a \textit{core} node;
\item If the TC degree of a node is no larger than the TC degrees of all of its
  neighbors, then the node is a \textit{margin} node;
\item If the number of neighbors with lower TC degrees equals to the number of
  neighbors with higher TC degrees, then the node is a \textit{bridge} node;
\item Otherwise, the node is a \textit{mediated} node.
\end{enumerate}

Let $\alpha={\#L(n)}/{\#N(n)}$ and $\beta={\#H(n)}/{\#N(n)}$, where $n$ is a
node, $\#L(n)$ is the number of neighbor nodes of $n$ with TC degrees lower than
$n$, $\#H(n)$ is the number of neighbor nodes of $n$ with TC degrees higher than
$n$, and $N(n)$ is the neighbors of $n$, then role of $n$ is distinguished by
\[
role(n)= \left\{
  \begin{array}{ll}
    core~node & \alpha > threshold(core) \\
    margin~node & \alpha = 0 \\
    bridge~node & \alpha = \beta \\
    mediated~node & otherwise
  \end{array}
\right.
\]
Where $threshold (core) \in (0.5, 1]$ controls the number of core nodes.

\begin{figure}[!htb]
  \centering \resizebox{0.6\columnwidth}{!}{\includegraphics{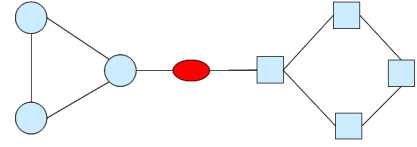}}
  \caption{The ellipse node is a topological center, and it is not a core node
    but a bridge node.}
  \label{fig:center-role}
\end{figure}

A node is a core node because it connects to more nodes or more important nodes.
A node is core node or not is decided by whether it has larger TC degrees than
its neighbors. However, the topological centers of a connected network may be
exceptions. In Fig. \ref{fig:node-roles}, node 2 is both the topological center
and a core node, but the ellipse node in Fig. \ref{fig:center-role} is the
topological center, and it is not a core node but a bridge node, although it has
higher TC degree than all of its neighbors. So it is significant to distinguish
the roles of topological centers. If the neighbors of a topological center are
all core nodes, then, the topological center is a bridge node, else the
topological center is a core node.

Researchers and papers may play such roles as \textit{source},
\textit{authority}, \textit{bee}, \textit{hub} and \textit{novice}
\cite{zhuge06:_discov_knowl_flow_scien}. The \textit{source},
\textit{authority}, and \textit{hub} may be \textit{core} nodes; \textit{bee}
nodes are often \textit{bridge} nodes; and the \textit{novice} may be the
\textit{margin} nodes or \textit{bridge} nodes.

In research network, a research group's leader usually has more publications and
cooperators. Correspondingly, they have more \textit{coauthor} relations
connecting to other researchers in the coauthor network. If each research group
is regarded as a community, the research group's leaders are the core nodes. The
fresh students have few publications and cooperators, so they are the margin
nodes in coauthor network. Visiting researchers and newly employed researchers
are bridge nodes, because they have cooperators in different research
communities. After the core nodes, the margin nodes and bridge nodes are
distinguished, the left nodes are mediated nodes. Usually, mediated nodes only
belong to one community.

In citation network, core nodes are the authority or hub papers having more
citations than others; the margin nodes are the novice papers or newly published
papers; and the bridge nodes connect two or more paper clusters. Each paper
cluster may belong to a specific research topic or discipline.

Funding decision-making and research promotion need to evaluate researchers and
their papers. Topological centrality can help distinguish the roles of
researchers and papers, and the roles can be used to evaluate researchers and
papers. TC degrees in the coauthor network help evaluate researchers, while TC
degrees in citation network help evaluate papers.

In research network, roles of nodes will change year by year. In the coauthor
network, a novice researcher may become an authority, a hub or even a bridge.
With more papers published, the TC degree of a node in a coauthor network will
become higher than its neighbors, and then the researcher become an authority or
hub. Cooperating with researchers in different research groups or even different
communities, a researcher becomes a bridge.

\subsection{Discovering Communities by Roles}

Tree in Fig. \ref{fig:centrality-compare-case} can be a coauthor network or a
citation network with directions of edges ignored. General community discovery
algorithms like GN algorithm cannot discover its communities, because the
betweenness of each edge is the same, and there is no way to choose the proper
edge for deletion. However, nodes in the coauthor networks and citation networks
play different roles, and communities can be discovered according to the roles
of nodes.

The roles of nodes can be used to discover communities. One way is to find the
core nodes, and then assign non-core nodes to the proper core nodes to form
communities. Algorithm \ref{alg:find-community} discovers communities by finding
core nodes for each non-core node.

\begin{algorithm}[!htb]%
  \centering
  \caption{Finding $k$ communities by core nodes}
  \label{alg:find-community}
  \begin{algorithmic}[1]%
    \REQUIRE a network $C$;

    \STATE Calculate the topological centrality degrees of nodes and links;
    \STATE Distinguish roles of nodes and add the core nodes into $CoreSet$; \FOR
    {node $x \in CoreSet$} \STATE $nodes(x) \leftarrow {x}$
    \ENDFOR

    \FOR {each non-core node $x$} \STATE Choose the nearest core nodes into
    $CandidateSet$ as the candidate nodes; \FOR {node $y \in CandidateSet$}
    \STATE $nodes(y) \leftarrow nodes(y) \cup x$;
    \ENDFOR
    \ENDFOR

    \WHILE{$|CoreSet| > k$} \STATE Merge two most tightly connected communities;
    \ENDWHILE

    \RETURN $k$ communities.
  \end{algorithmic}
\end{algorithm}

The time complexity of algorithm \ref{alg:find-community} is $O(n(n + m))$. The
number of core nodes can be controlled by setting the threshold of
$\#L(n)/\#A(n)$. If there are more than one candidate core nodes, then the node
should be classified into different communities, and the bridge nodes are often
classified into several communities at the same time.

This way can discover communities globally in a network. If the number of
communities is too many, the closely connected communities can be merged into
larger communities. Closely connected communities may share many nodes and
links, or there are many external connections between them. Suppose the number
of communities is $k$, Algorithm \ref{alg:merge-community} merges communities.

\begin{algorithm}[!htb]%
  \centering
  \caption{Merging communities}
  \label{alg:merge-community}
  \begin{algorithmic}[1]%
    \REQUIRE the number of communities $k$;
    
    \STATE Step 1. If the number of communities is less than $k$, then goto Step
    4. \STATE Step 2. Calculate the Jaccard similarity of node sets of each
    community pair. Suppose $A$ and $B$ are two communities, Jaccard similarity
    of $A$ and $B$ is calculated by $$Jaccard(A,B) =\frac{|A \cap B|}{|A \cup
      B|}. $$ If all the Jaccard similarities of community pairs equal to $0$,
    then goto Step 3; else, find the community pairs have the largest Jaccard
    similarity, and merge them into a larger community respectively. Goto Step
    1. \STATE Step 3. Count the external links between community pairs. An
    external link has two end nodes in two different communities respectively.
    If all the numbers of external link set equal to $0$, then goto Step 4;
    else, find the community pairs have the maximum external links, and merge
    them into a larger community respectively. Goto Step 1. \STATE Step 4. Stop
    merging communities.

\end{algorithmic}
\end{algorithm}

Another way is to find the core nodes first, and then expand from a node to form
local communities. According to role of nodes, the community expansion needs to
consider the following cases.

\begin{enumerate}
\item Forming local community according to core node. Algorithm
  \ref{alg:expand-community} is for discovering local communities from a core
  node. A community may have more than one core node. If two communities share
  many common nodes and links, then the two communities can be merged into a
  larger community. This way can find the research groups in a coauthor network,
  and can find the specific topic related paper clusters in the citation
  network.

\item Form local community according to non-core node. To find local communities
  from a non-core node, it is necessary to find the core nodes connected to the
  node. Before finding communities of a non-core node, all the core nodes in the
  network should be found first. Then expand the local communities from the
  nearest core nodes connected to the non-core node respectively.

\item Finding local community of a set of nodes. Given a set of nodes, the local
  community can be found as follows.

  \begin{enumerate}%
  \item For each node, find the core nodes connected to it until the topological
    center is found; all the core nodes are added to \textit{coreSet}.
  \item Building the subgraph containing these nodes and nodes in
    \textit{coreSet}; and,
  \item Expanding the local community from the nodes in \textit{coreSet}.
  \end{enumerate}

\end{enumerate}

  \begin{algorithm}[!htb]%
    \centering
    \caption{Expanding community from a core node}
    \label{alg:expand-community}
    \begin{algorithmic}[1]%
      \REQUIRE A core node $c$ and a connected network $G$; \STATE
      $nodeQueue \leftarrow \{c\}$, $nodeSet \leftarrow \{c\}$, $linkSet \leftarrow \{\}$;

      \WHILE {$nodeQueue \ne \{\}$} \STATE Fetch a node $x$ from $nodeQueue$;
      \FOR { $y$ is the neighbor node of $x$} \STATE Distinguish the role of
      $y$; \IF{ $(y \notin nodeSet)$ and ($y$ is not a core node) and
        ($nodeWeight(y) < nodeWeight(x)$)} \STATE $nodeQueue \leftarrow nodeQueue \cup y$;
      \STATE $nodeSet \leftarrow nodeSet \cup y$; \STATE $linkSet \leftarrow linkSet \cup link(x,y)$;
      \ENDIF
      \ENDFOR
      \ENDWHILE
      \RETURN $linkSet$.
    \end{algorithmic}
  \end{algorithm}

\begin{figure}[!htb]
  \centering \resizebox{0.7\columnwidth}{!}{\includegraphics{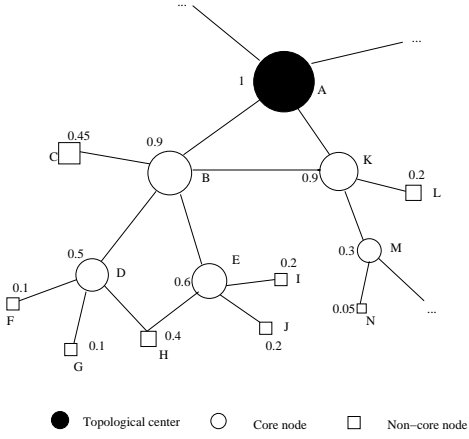}}
  \caption{A simple case for finding community: circle nodes are core nodes;
    square nodes are non-core nodes.}
  \label{fig:community-case}
\end{figure}

  Fig. \ref{fig:community-case} shows a segment of network with TC degrees of
  nodes. We can find a local community from a core node, a non-core node, and a
  set of nodes as follows.

\begin{enumerate}
\item Finding local community of core node $B$. The process is shown as Table
  \ref{tab:local-community-core}.

\begin{table}[!htb]
  \centering
  \caption{Finding local community of core node $B$}
  \label{tab:local-community-core}
  \begin{tabular}{|l|l|l|l|l|}
    \hline
    Step  & Node  & nodeQueue  & nodeSet  & Expanded 
    \\ \hline \hline
    0  & B  & B  & B  & C, D, E \\ \hline 
    1  & C  & D, E  & B, C  & \\ \hline 
    2  & D  & E  & B, C, D  & F, G, H \\ \hline 
    3  & E  & F, G, H  & B, C, D, E  & I, J \\ \hline 
    4  & F  & G, H, I, J  & B, C, D, E, F  & \\ \hline 
    5  & G  & H, I, J  & B, C, D, E, F, G  & \\ \hline 
    6  & H  & I, J  & B, C, D, E, F, G, H  & \\ \hline 
    7  & I  & J  & B, C, D, E, F, G, H, I  & \\ \hline 
    8  & J &  & B, C, D, E, F, G, H, I, J & \\ \hline 
  \end{tabular}
\end{table}

\item Finding local community of non-core node $F$ is to find the nearest core
  node $D$, then find the local community from $D$. The expansion process is
  shown in Table \ref{tab:local-community-noncore}.

\begin{table}[!htb]
  \centering
  \caption{Finding local community of non-core node $F$}
  \label{tab:local-community-noncore}
  \begin{tabular}{|l|l|l|l|l|}
    \hline
    Step  & Node  & nodeQueue  & nodeSet  & Expanded 
    \\ \hline \hline
    0  & D  & D  & D, F  & G, H \\ \hline 
    1  & G  & H  & D, F, G  & \\ \hline 
    2  & H  &  & D, F, G, H  & \\ \hline 
  \end{tabular}
\end{table}

\item Finding local community of a node set $\{D, I, J\}$. $D$ is a core node,
  while $I$ and $J$ are two non-core nodes. If $D$ is the core node of the
  community containing $I$ and $J$, then $\{D, I, J\}$ forms the local
  community. However, $D$ is not the core node of the community containing $I$
  and $J$. The possible core nodes of the community containing $D$ are $\{D, B,
  A\}$; the possible core nodes of the community containing $I$ and $J$ are the
  same, that is, $\{E, B, A\}$. Then, we can construct the subgraph containing
  node $D$, $I$ and $J$ and their possible core nodes $D$, $E$, $B$ and $A$ as
  shown in Fig. \ref{fig:subfig-community}.

\begin{figure}[!htb]
  \centering
  \resizebox{0.4\columnwidth}{!}{\includegraphics{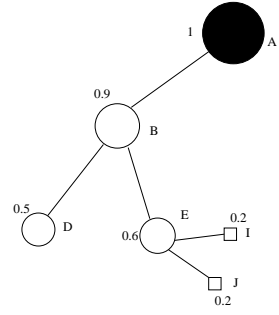}}
  \caption{Subgraph containing node D, I, J and the possible core nodes D, E, B
    and A.}
  \label{fig:subfig-community}
\end{figure}

From the subgraph, we know that $B$ the nearest core node of the community
containing $D$, $I$ and $J$. Then, we can expand from $B$ to find the local
community containing node $D$, $I$ and $J$ as mentioned in case (1).

\end{enumerate}

In research network, this way can find research team members of a researcher in
a coauthor network and find topic-related papers of a paper in a citation
network.

Given a set of papers, the coauthor relations form the coauthor network, and the
citation relations form the citation network. After the TC degrees are
calculated, the research groups can be discovered, and the papers can be
clustered by citation relations. Researchers in the same communities may share
the similar research interests, while papers in the same clusters are topic
related. Topic-related papers can be recommended to researchers having similar
research interests. Global communities show research groups and research topics
in the paper set, while the local community expansion way help recommend papers
in a large paper set to appropriate readers.

When making a funding decision, it is necessary to evaluate the status of a
research group, cooperators, and publications. The discovered communities in
coauthor network show the research groups of a research area, while the
discovered communities in citation network show paper clusters in the research
area. And, the roles of the researcher and his/her publications can be
distinguished by TC degrees.

\section{Application: Discovering Backbone in Research Network}

Given a set of research papers, research networks such as coauthor networks and
citation networks can be constructed. Metadata of papers in computer science are
often stored in Bibtex or XML files provided by online digital libraries such as
Google Scholar, ACM Portal (http://portal.acm.org), IEEE digital library
(http://ieeexplore.ieee.org), DBLP
(http://www.informatik.uni-trier.de/\~{}ley/db/) and Citeseer
(http://citeseer.ist.psu.edu) etc.

\subsection{Structures of Research Network}

Researchers and the \textit{coauthor} relation form the coauthor network.
Coauthors of a paper formulate the motif \cite{milo2002nms} of research network.
A coauthor relation from $A$ to $B$ means that $A$ and $B$ are coauthors of the
same paper, and $A$ is before $B$ in the author list.

\begin{figure}[!htb]
  \centering
  \resizebox{0.9\columnwidth}{!}{\includegraphics{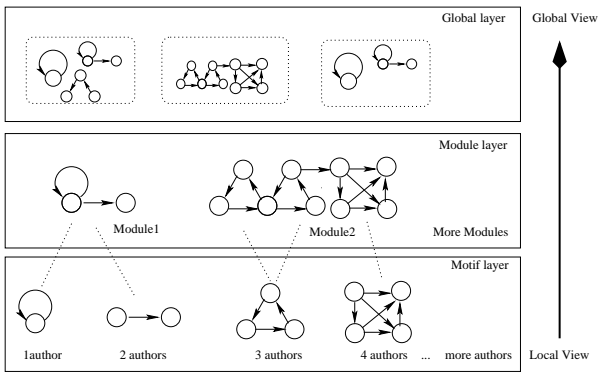}}
  \caption{Structure of coauthor network from local view to the global view: (1)
    the bottom layer contains the motifs; (2) the middle layer contains the
    modules combing one or more motifs; (3) the top layer contains the networks
    of modules.}
  \label{fig:coauthor-structure}
\end{figure}

Fig. \ref{fig:coauthor-structure} shows the structure of the coauthor network.
With the directions of coauthor relations ignored, each motif describes the
cooperation between authors of a paper: a loop for the sole author, an edge
between two authors, a triangle for three authors, and a complete graph for $n
(n > 3)$ authors. Coauthor network has three layers from local view to the
global view: \textit{motif layer}, \textit{module layer} and \textit{global
  layer}. Nodes' degrees in coauthor network reflect the active degrees of
researchers. The in-links reflect the hub characteristic, while out-links
reflect authority.

\begin{figure}[!htb]
  \resizebox{\columnwidth}{!}{\includegraphics{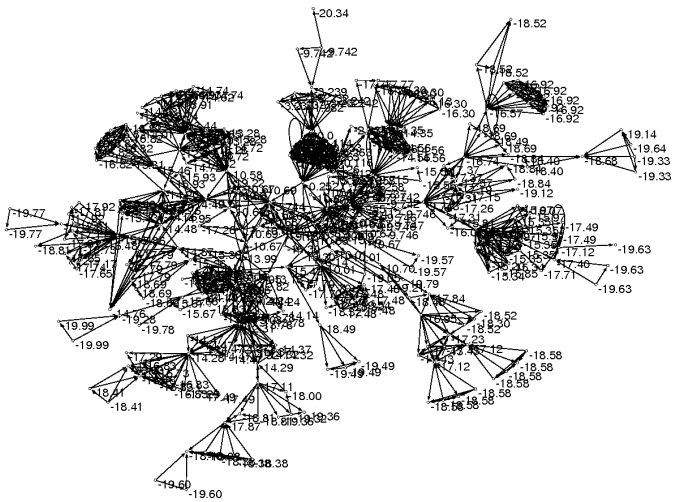}}
  \caption{The largest module of coauthor networks of ISWC data set with
    weights: the topological centrality degrees are transformed by function
    $log()$.}
  \label{fig:tc-largest-coauthor}
\end{figure}

Our first dataset collects papers of the International Semantic Web Conference
(ISWC) from $2002$ to $2007$. The number of researchers and papers are $935$ and
$401$ respectively. The number of \textit{coauthor} relations is $2286$. The
number of citation relationship is $236$, and citation relations are considered
between the paper pairs both in ISWC. The number of \textit{authorOf} relations
is $1362$. Fig. \ref{fig:tc-largest-coauthor} shows the node TC degrees of the
largest module of coauthor networks with a circular layout. The central nodes
have higher TC degrees, and the topological nodes have the highest centrality
$1$. From a topological center to the margins, the TC degrees reduce to $0$ step
by step. If the number of nodes are very huge, the TC degrees are very small,
and function $log()$ maps the TC from interval $(0, 1]$ to $(-21, 0]$, and the
order of node TC keeps unchanging.

\begin{figure}[!htb]
  \resizebox{0.95\columnwidth}{!}{\includegraphics{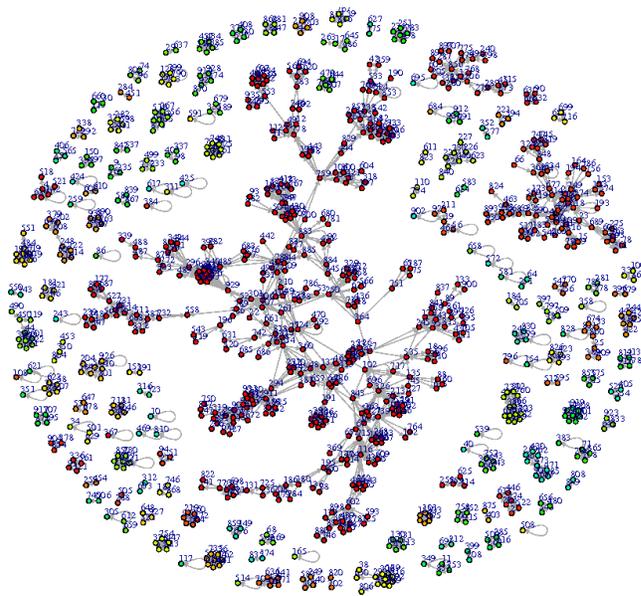}}
  \caption{Coauthor networks of ISWC data set: $147$ modules, $935$ researchers
    and $2286$ coauthor relations.}
  \label{fig:coauthor-iswc}
\end{figure}

\begin{figure}[!htb]
  \resizebox{0.95\columnwidth}{!}{\includegraphics{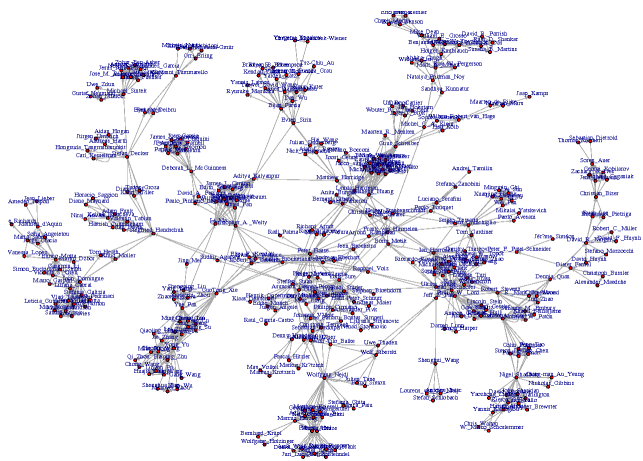}}
  \caption{The largest module of coauthor networks of ISWC data set.}
  \label{fig:coauthor-name-iswc}
\end{figure}

  Fig. \ref{fig:coauthor-iswc} shows the modules in coauthor network of ISWC
  dataset. It contains $147$ modules, $935$ researchers and $2286$ coauthor
  relations. Fig. \ref{fig:coauthor-name-iswc} shows the largest module of Fig.
  \ref{fig:coauthor-iswc}. It contains $370$ researchers and $1227$ coauthor
  relations.

The number of coauthor relations between two researchers reflects the frequency
of their cooperation. Node degrees in coauthor network reflect the active
degrees of researchers. The in-links reflect the hub characteristic, while
out-links reflect authority.

The density of a module is reflected by the frequency of cooperation between
researchers. The average cooperation active degree between each pair of
researchers, called cooperation density, can be used to assess the active degree
of a research community. Cooperation density is the number of coauthor relations
dividing the number of researchers.

\textbf{Theorem 1.} A module $M$ of coauthor network has $n$ researchers, the
lower bound and upper bound of module density are within the range $[(n-1)/n, n
- 1]$.

\textit{Proof.} Suppose $M$ is a connected digraph with $n$ nodes. The lower
bound of density: the number of edges is $n - 1$ at least, otherwise there will
be some isolated researchers. So the lower bound density of $M$ is $(n-1)/n$.
The upper bound of density: if there are at most one directed edge between two
nodes, then the number of edges in $M$ is $n(n - 1)$ at most. So the upper bound
density of $M$ is $n - 1$. Therefore, the lower bound and upper bound of module
density of module $M$ with $n$ nodes are within the range $[(n-1)/n, n-1]$.
$\Box$

\begin{figure}[!htb]
  \centering
  \begin{tabular}{c}
    \resizebox{0.6\columnwidth}{!}{\includegraphics{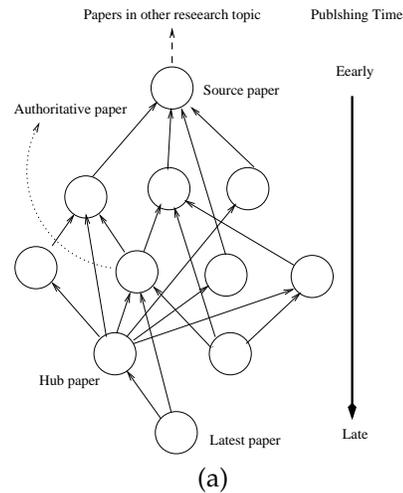}} \\
    (a)  \\ \\
    \resizebox{0.7\columnwidth}{!}{\includegraphics{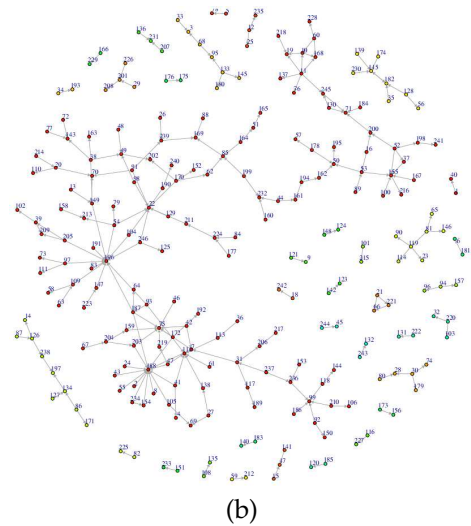}} \\
    (b)  \\
  \end{tabular}
  \caption{Structure and instance of citation network: (a) The structure of
    citation network. (b) The citation network of ISWC data set.}
  \label{fig:citation-network}
\end{figure}

Citation network is a directed acyclic graph (DAG). Each paper has the fixed
publishing time, and papers can only cite the papers already published, so there
are no cycles in the citation network. Citation is direction sensitive, and it
implies the time sequential relationship between two papers. Fig.
\ref{fig:citation-network}a shows a module of citation network. Papers in the
same module are topic related.

Citation relations show the relevance between research papers, and paper
communities can be discovered by citation relations. Citations in the community
show the relevance between papers, while citations between paper communities
show the relevance of research topics.

Fig. \ref{fig:citation-network}b shows the modules in citation network of ISWC
dataset. It contains $36$ modules, and the largest module contains $142$ papers
and $165$ citation relations. All the citation relations are between papers
published in ISWC. The connectivity density is less than the connectivity
density of coauthor network. The citation density of a module reflects the
relevance between the papers. The citation density is the number of citations
dividing the number of papers.

\subsection{Topological Centrality based Backbone Network}

In a network, after roles of nodes are distinguished by the node TC degrees,
core nodes and edges among them form a subgraph, called \textit{backbone
  network}. The end nodes of edges in the backbone network are both core nodes.

\begin{figure}[!htb]
  \centering
  \resizebox{0.7\columnwidth}{!}{\includegraphics{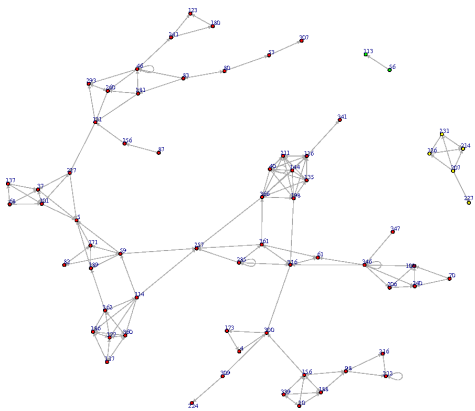}}
  \caption{Backbone network of the largest module of coauthor network of ISWC
    Dataset from $2002$ to $2007$.}
  \label{fig:coauthor-backbone-iswc}
\end{figure}

The backbone network consists of core nodes. It is useful for visualization and
browsing, and can play the following roles in scientific research:

\begin{enumerate}%
\item It helps display the research network of different levels. Each community
  can be represented by the core nodes in the backbone network. When a core node
  is focused, the detailed information of its local community can be browsed.
  
\item It shows the important researchers in a coauthor network. When a research
  community or research group is mentioned, the leaders of the community or the
  head of the research group are well known. Fig.
  \ref{fig:coauthor-backbone-iswc} shows the backbone network of the largest
  module of the coauthor networks of ISWC data set. The threshold of the core
  nodes is $0.5$, and the threshold of the margin nodes is $0$. It contains all
  of the core nodes and the coauthor relations among them. Most of the core
  nodes are connected, and this verifies the ``rich club'' phenomenon
  \cite{colizza2006drc}: richer nodes are more possibly connected with other
  richer nodes. Some core nodes formulate the connected components alone,
  because the bridge nodes between them are non-core nodes.

\item Backbone network of coauthor network can be used to propagate information.
  Coauthor network is a kind of social network. Core nodes are important during
  the information propagation because they have more impact in their
  communities. Suppose an invitation of PC members needs to be sent, the
  researchers in the backbone network should take the priority.

\item Papers formulate communities via the citation relations, and papers in a
  community share the same or relevant research topics. Core nodes are often
  important papers citing or are cited by more important papers. The backbone
  network of citation network helps find the development and history of a
  research area or a research topic. Core nodes and its neighbors reflect the
  main achievements at different research stages.

\item Paper publication venue network contains conferences and journals. Other
  research resources such as researchers, papers and publishers connect
  conferences and journals into a connected network. To find the citations among
  conferences and journals, the sub-network containing conferences, journals and
  papers can be built. If a super node represents the conference or journal
  containing papers, then citation relations in the super nodes and between
  different super nodes can be counted. The number of external citations
  reflects the relevance of conferences and journals.

\end{enumerate}

Similarly, the relevance of publishers' businesses, projects, and institutions
can be analyzed. The relevance of publishers is reflected by the relevance
between books and papers published by them. The relevance among projects is
reflected by the cooperation between researchers taking part in the projects and
citations between papers supported by the projects. The relevance between
institutions can also be reflected by the relevance between researchers and
papers.

\subsection{Evolution of Backbone Networks}

Backbone networks can be used to study the development of scientific research.
Backbone networks sorted by years reflect the evolvement of research networks.
Similarly, the evolvement of backbone networks in citation network, paper venue
network, and institution networks etc can be studied.

\begin{figure}[!htb]
  \centering

  \begin{tabular}{lr}
  
    \resizebox{!}{0.1\textheight}{\includegraphics{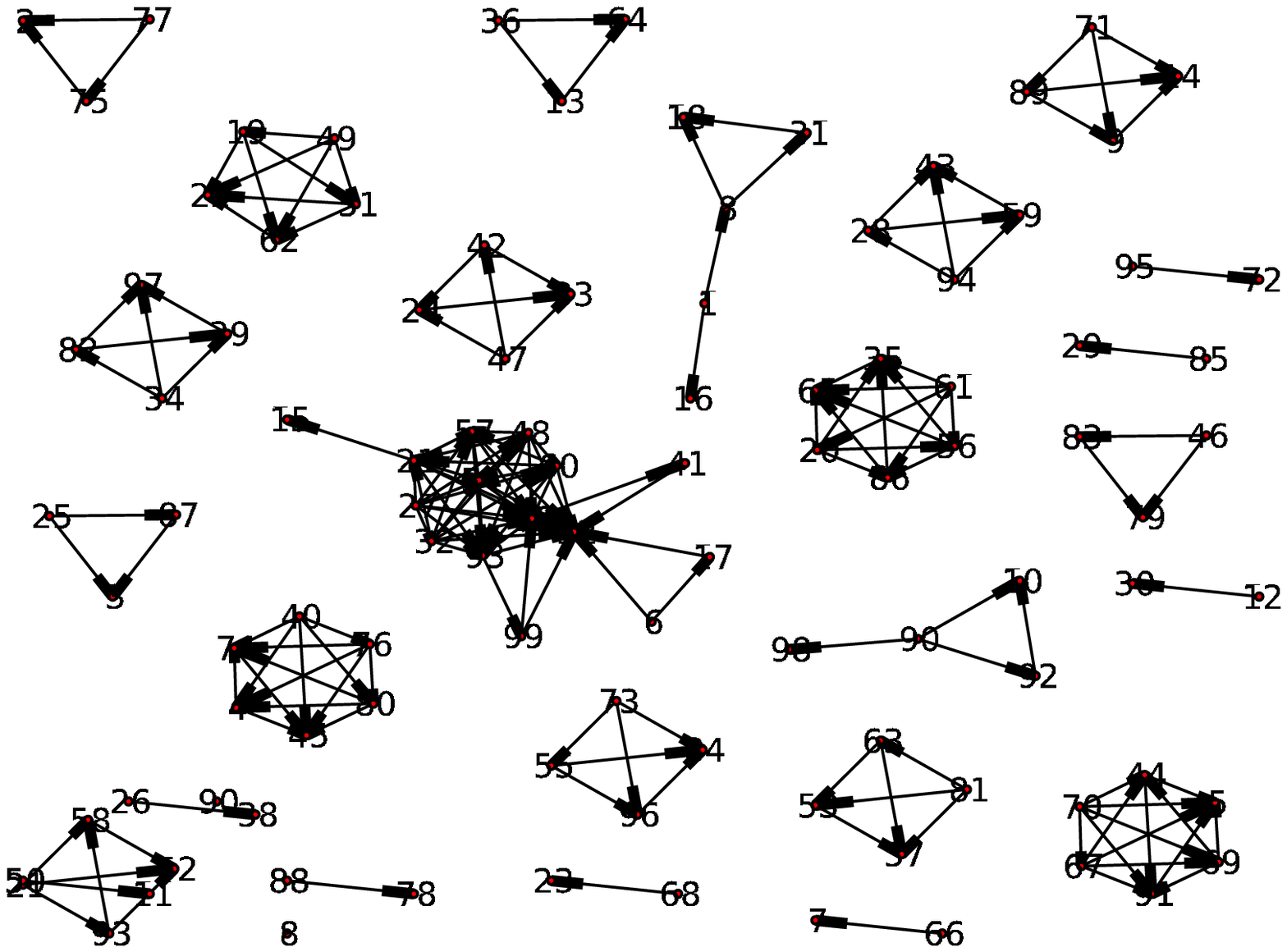}} & 
    \resizebox{!}{0.1\textheight}{\includegraphics{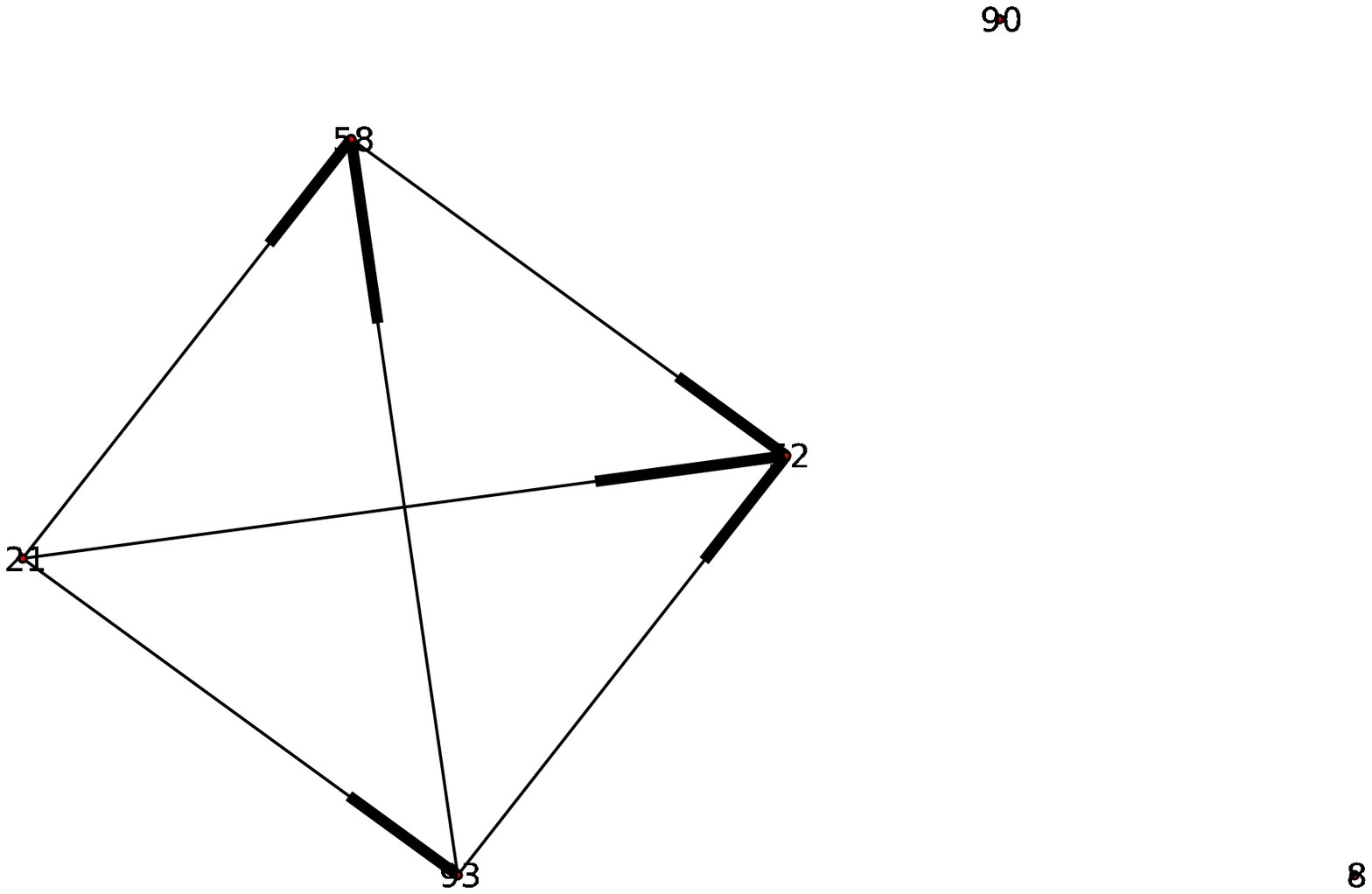}} \\ \\
  
    \resizebox{!}{0.1\textheight}{\includegraphics{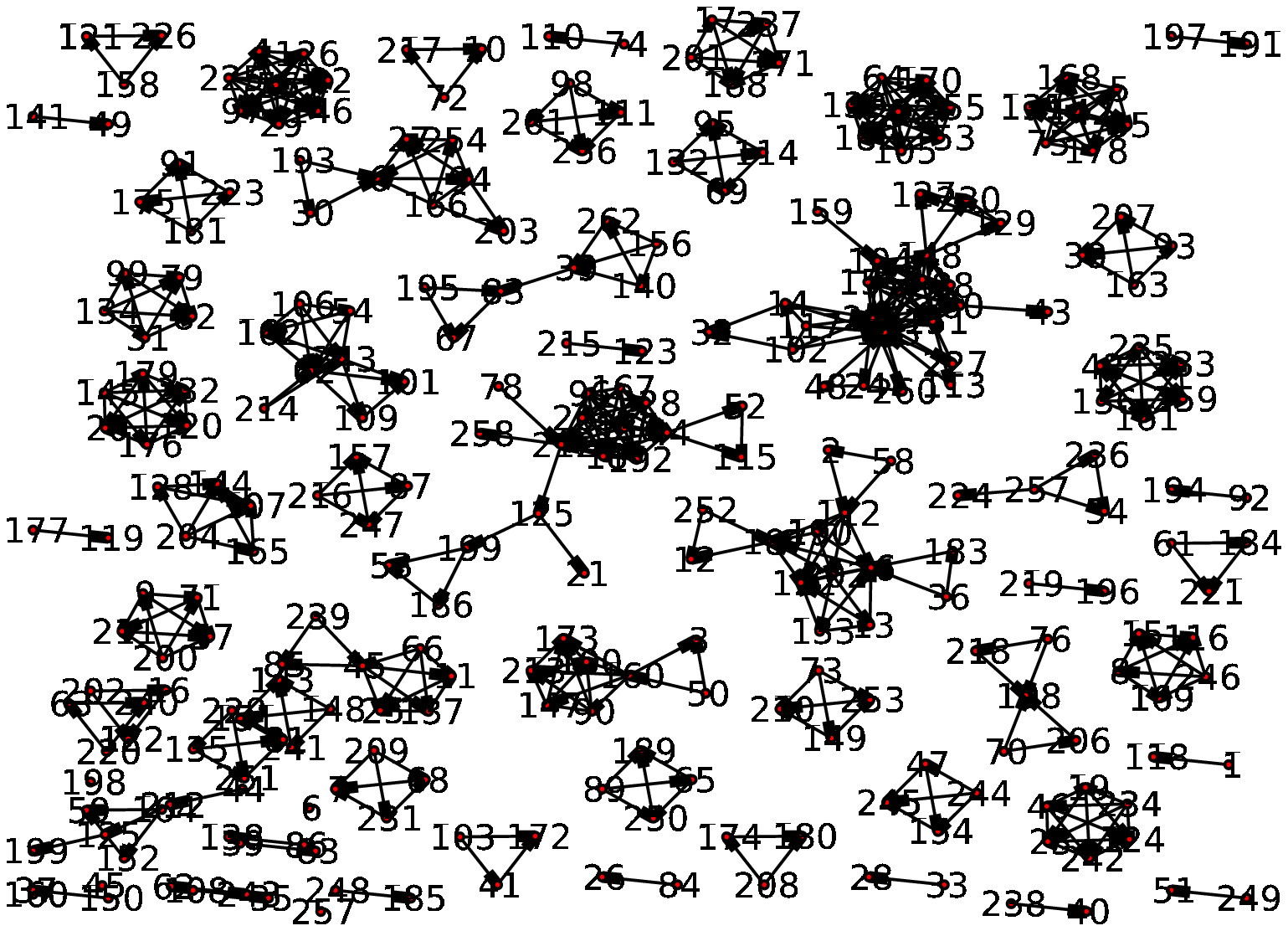}} & 
    \resizebox{!}{0.1\textheight}{\includegraphics{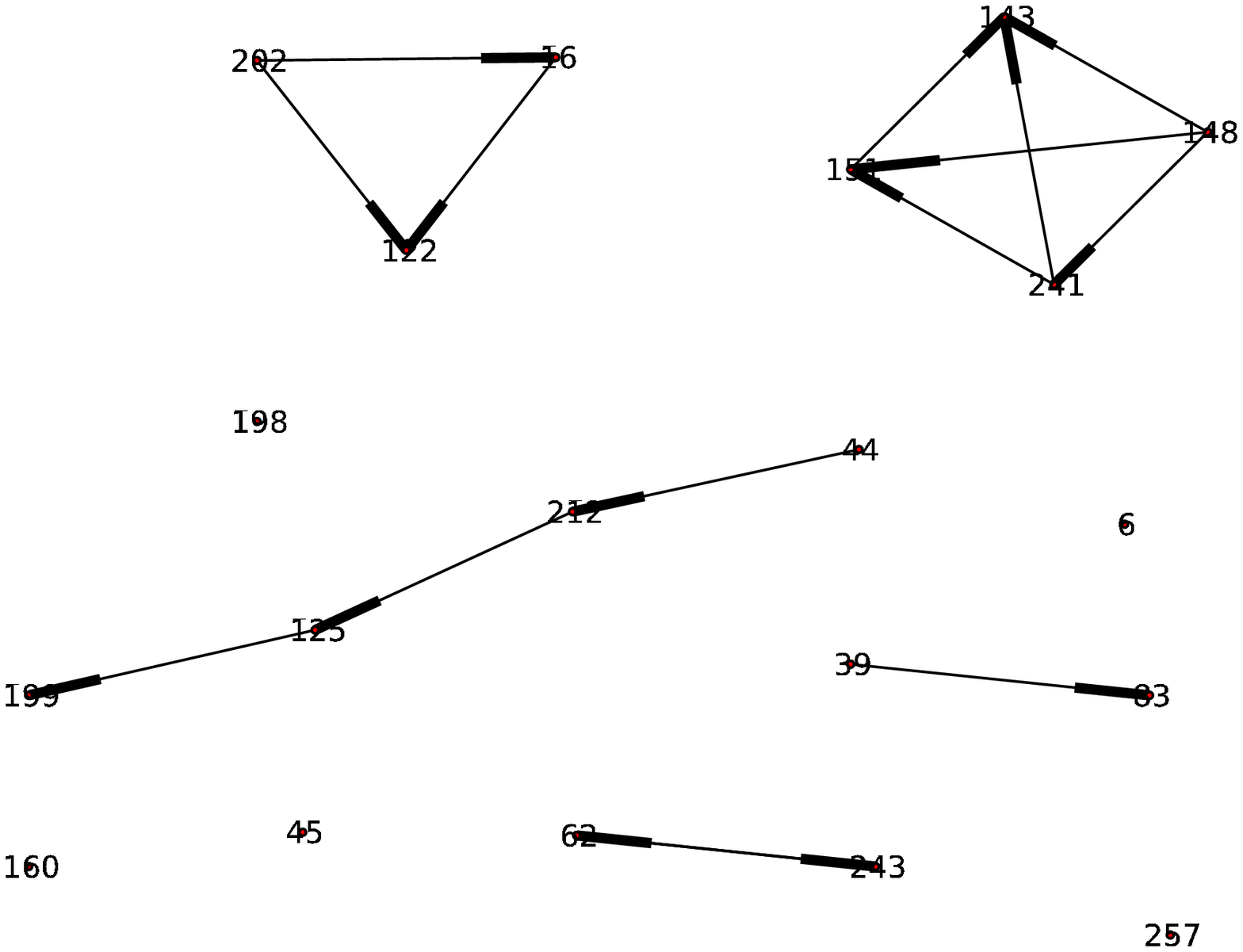}} \\ \\
  
    \resizebox{!}{0.1\textheight}{\includegraphics{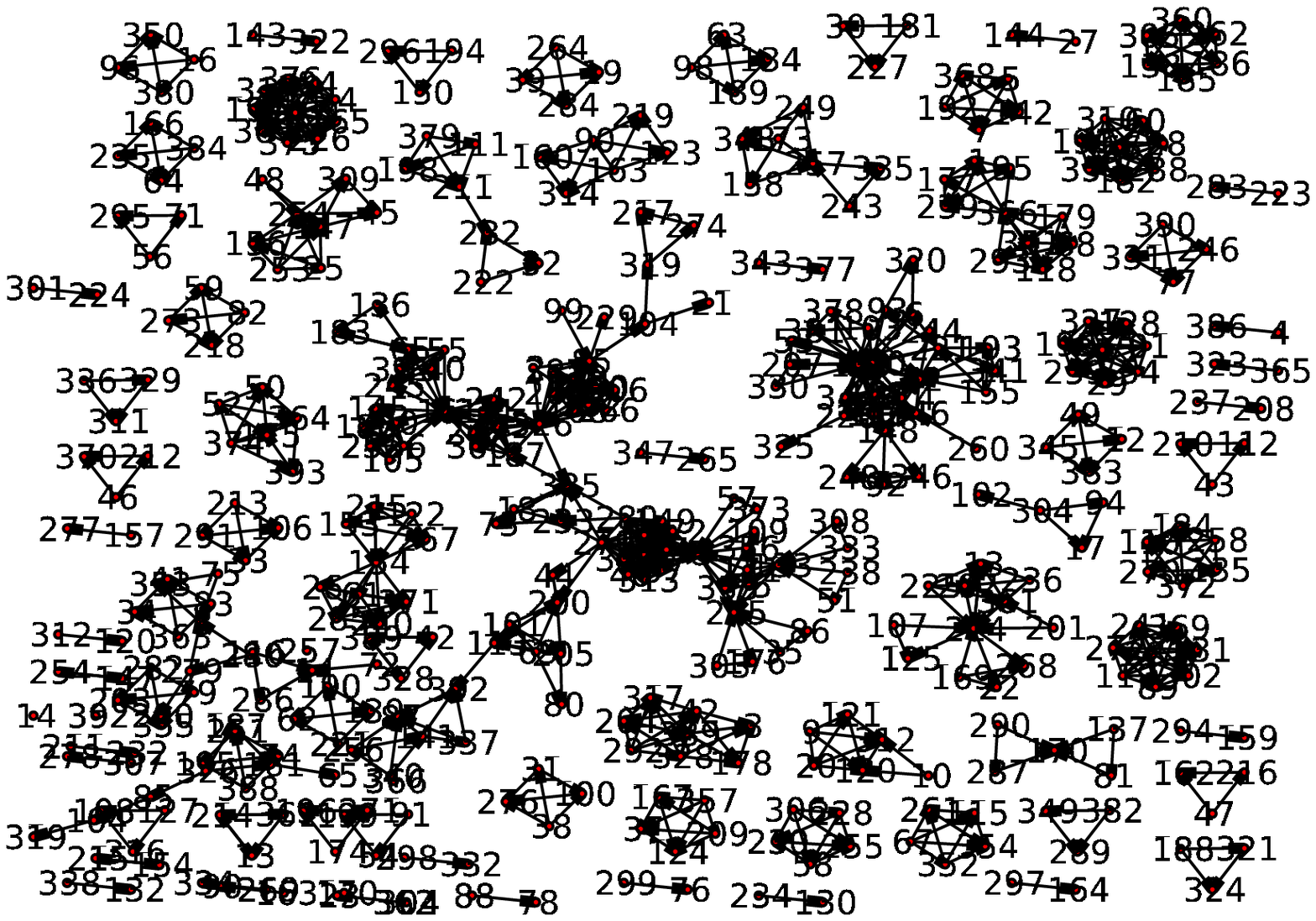}} & 
    \resizebox{!}{0.1\textheight}{\includegraphics{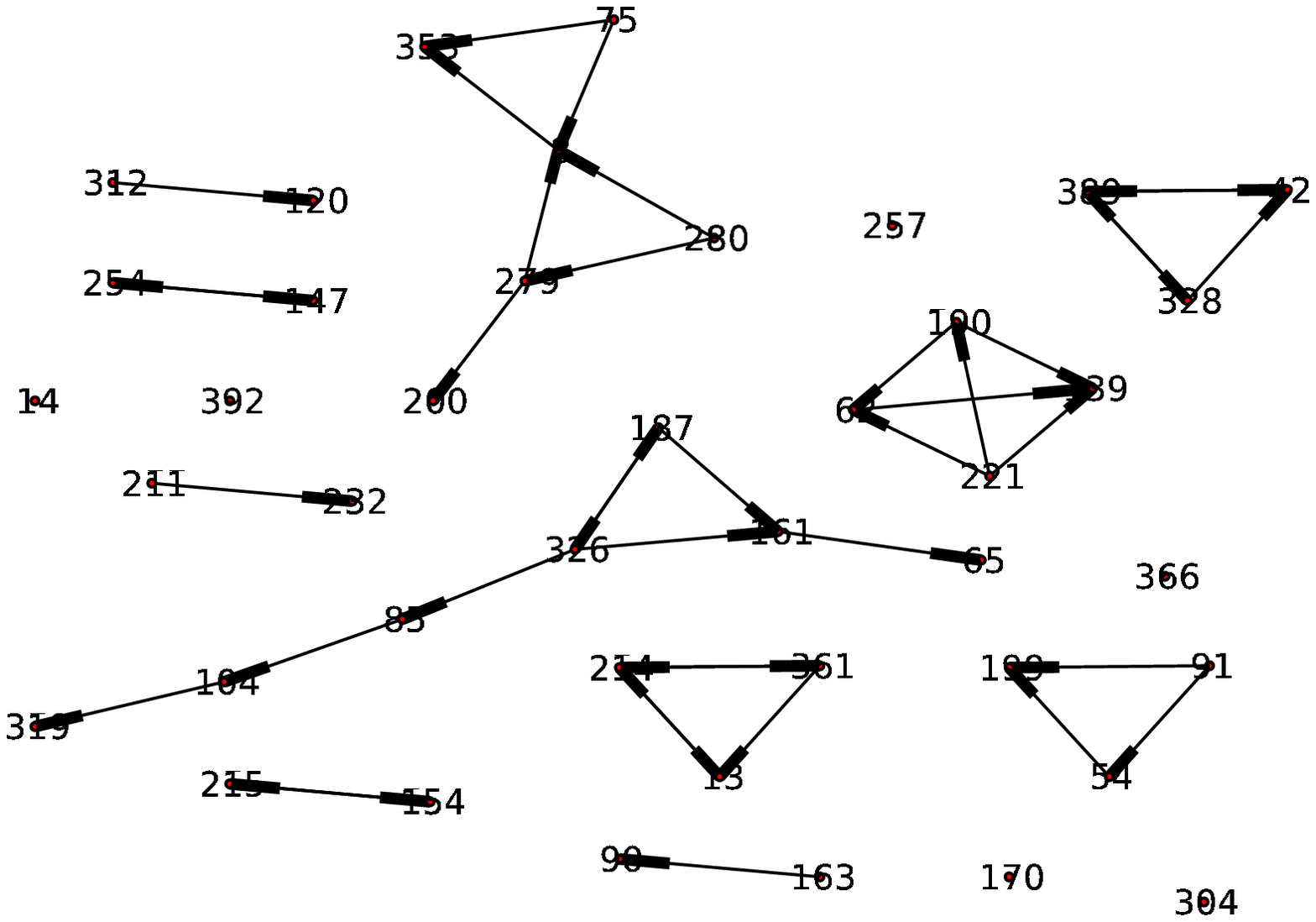}} \\ \\
  
    \resizebox{!}{0.1\textheight}{\includegraphics{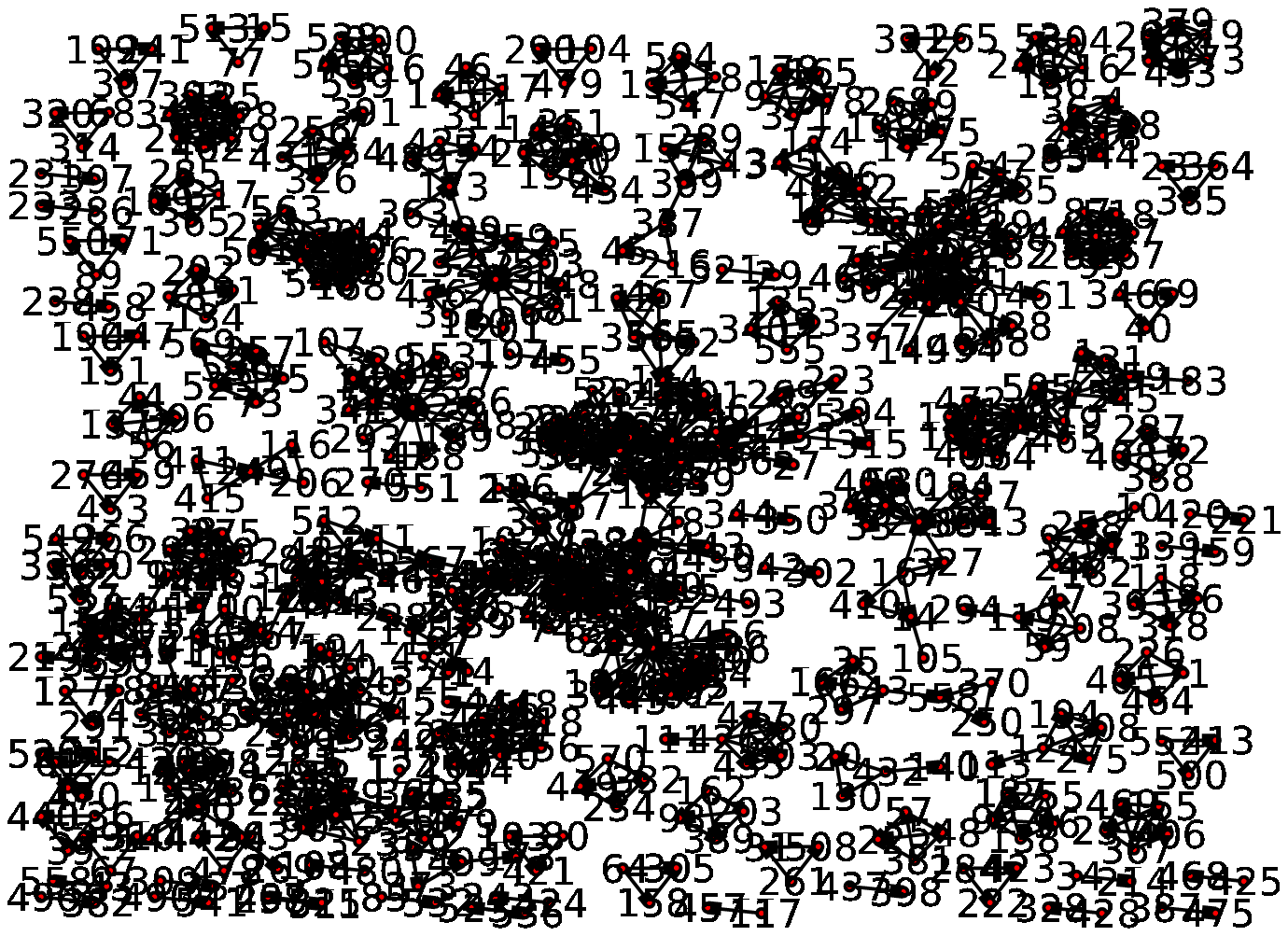}} & 
    \resizebox{!}{0.1\textheight}{\includegraphics{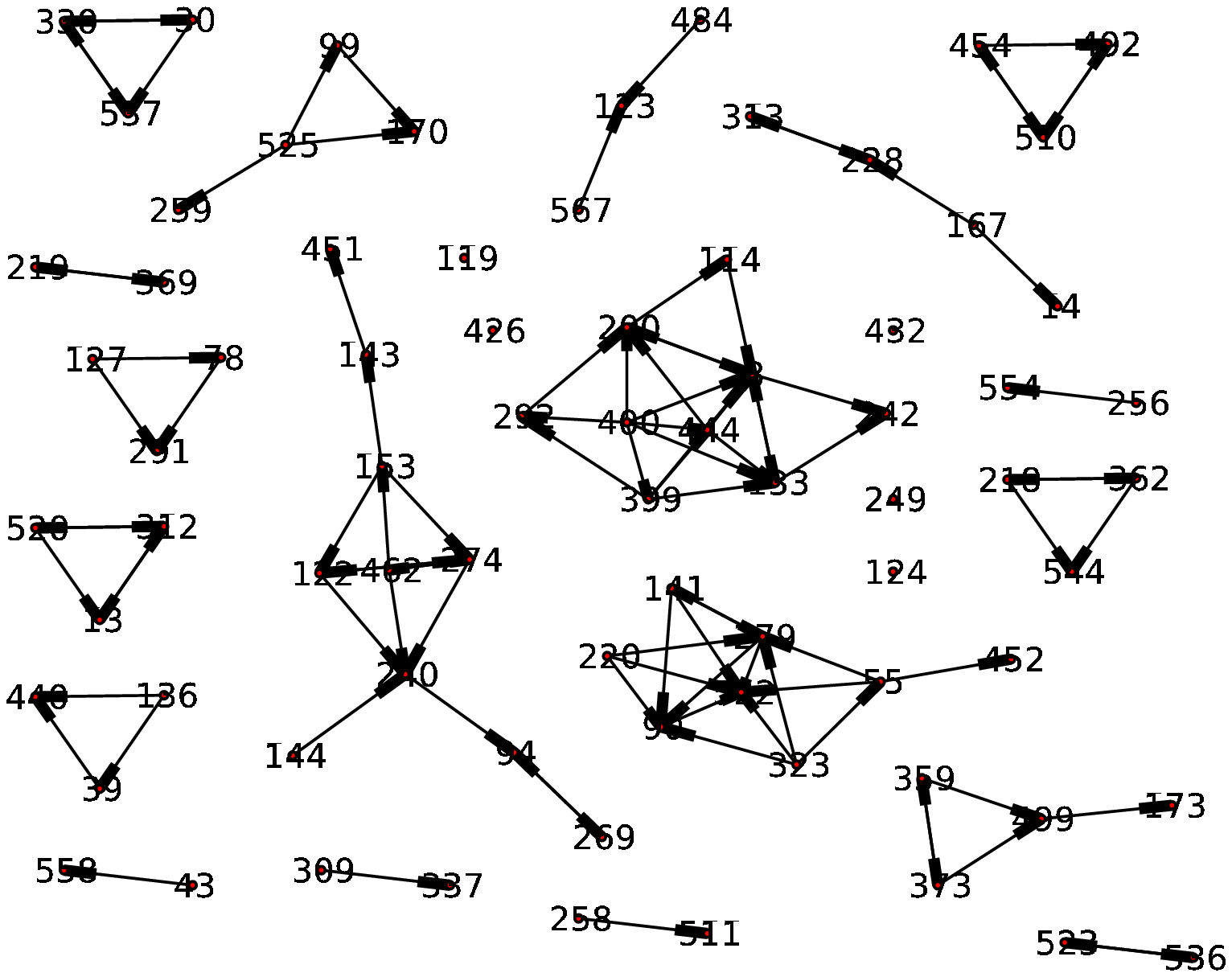}} \\ \\
  
    \resizebox{!}{0.1\textheight}{\includegraphics{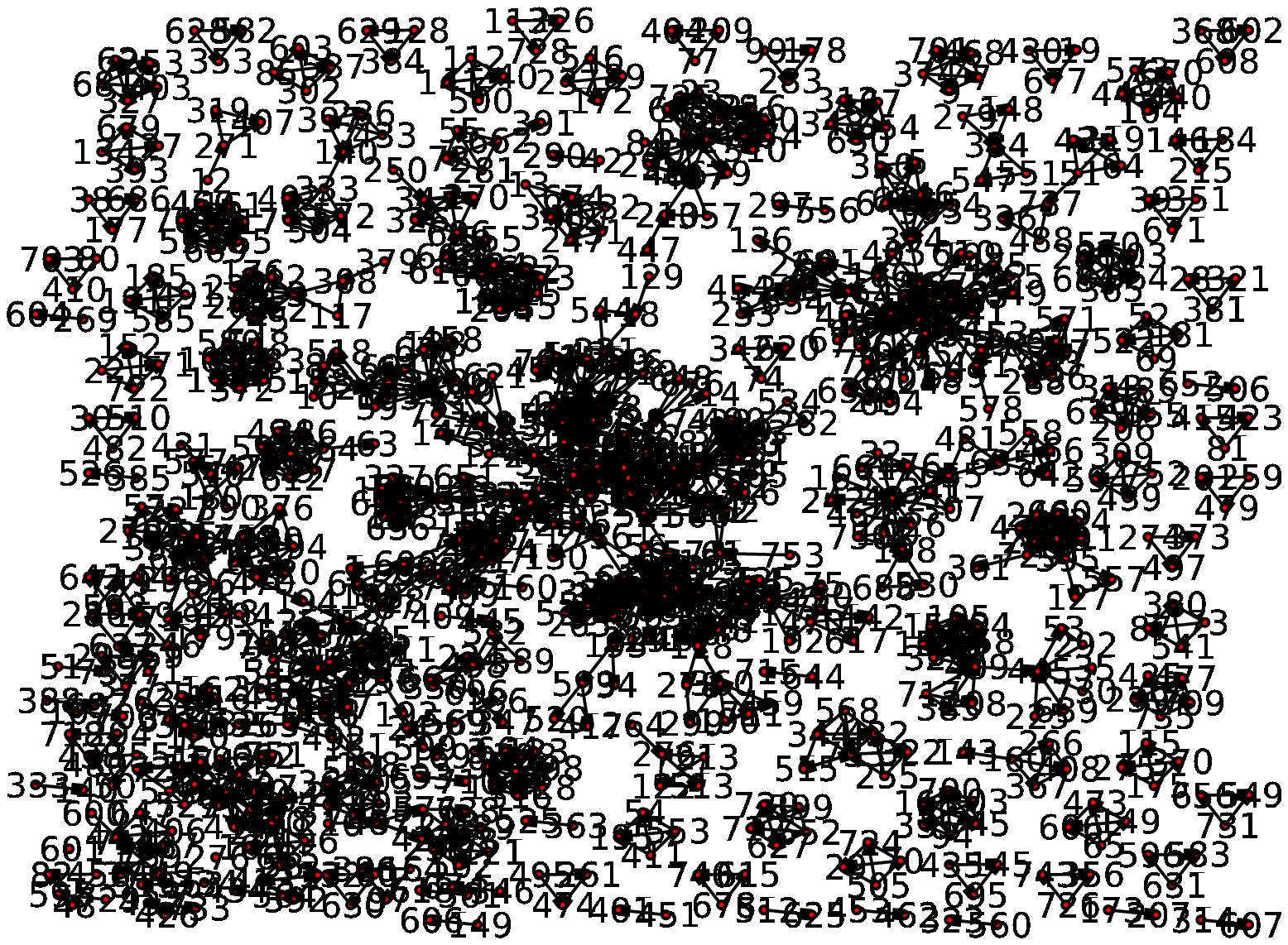}} & 
    \resizebox{!}{0.1\textheight}{\includegraphics{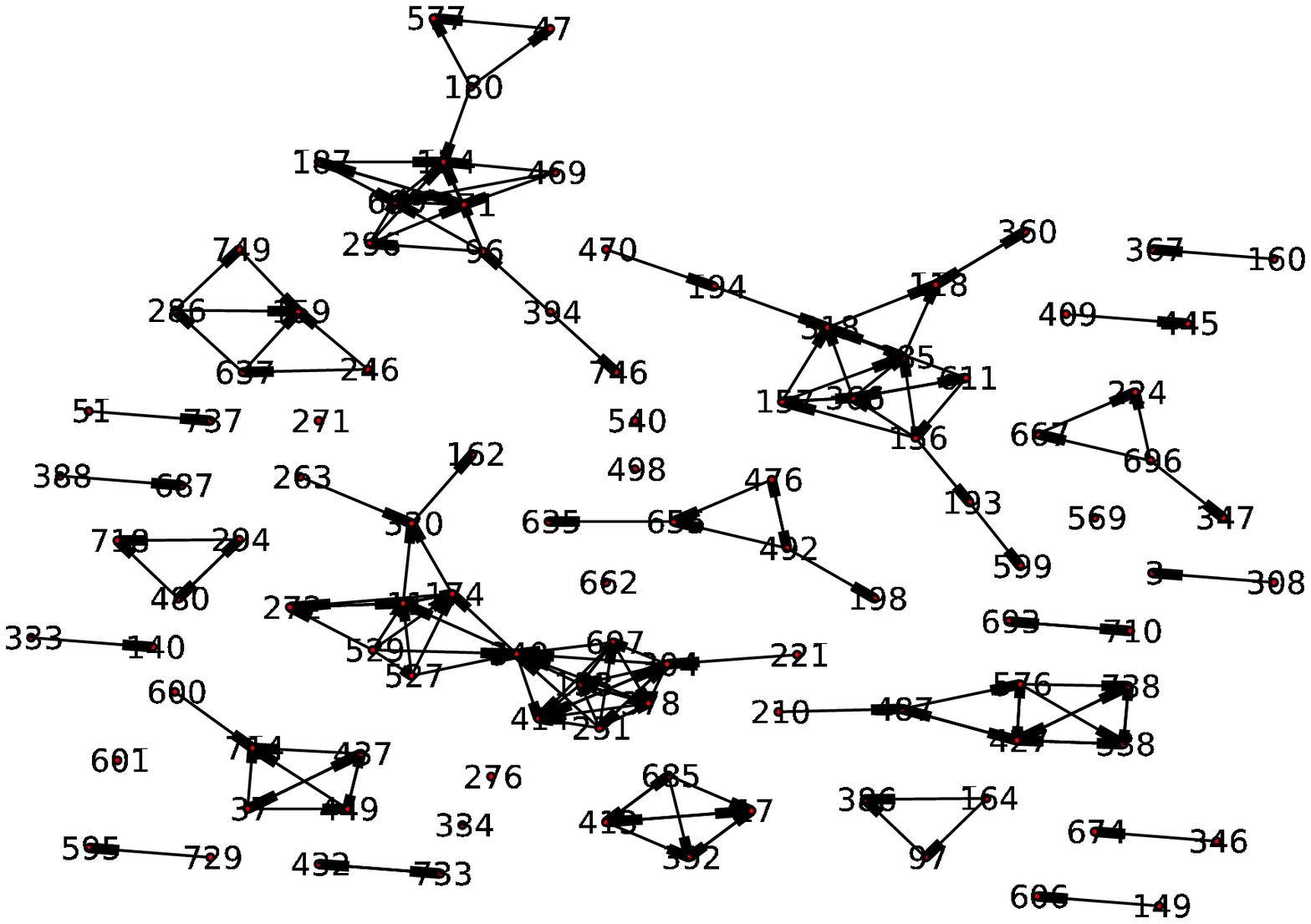}} \\ \\
  
    \resizebox{!}{0.11\textheight}{\includegraphics{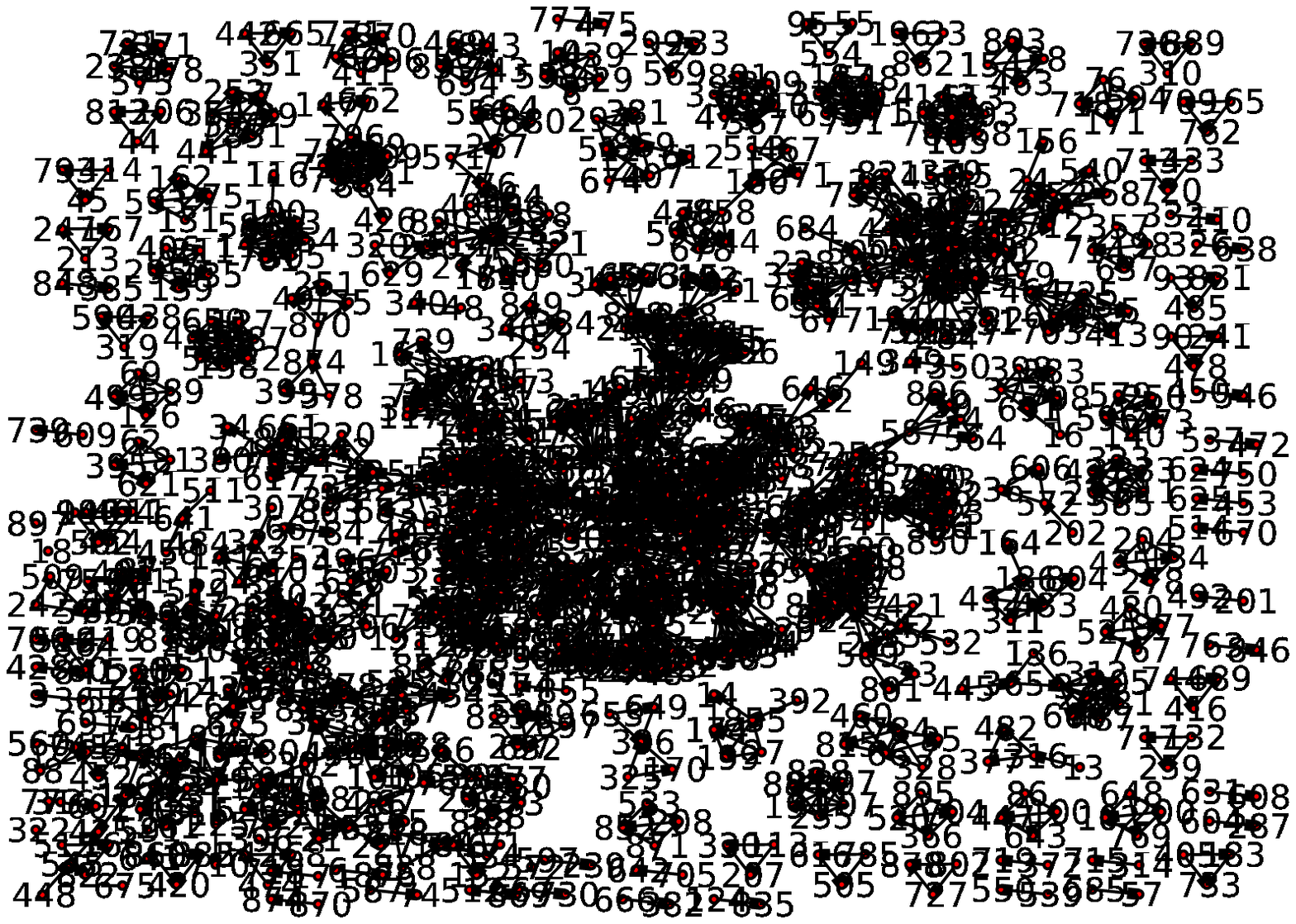}} & 
    \resizebox{!}{0.11\textheight}{\includegraphics{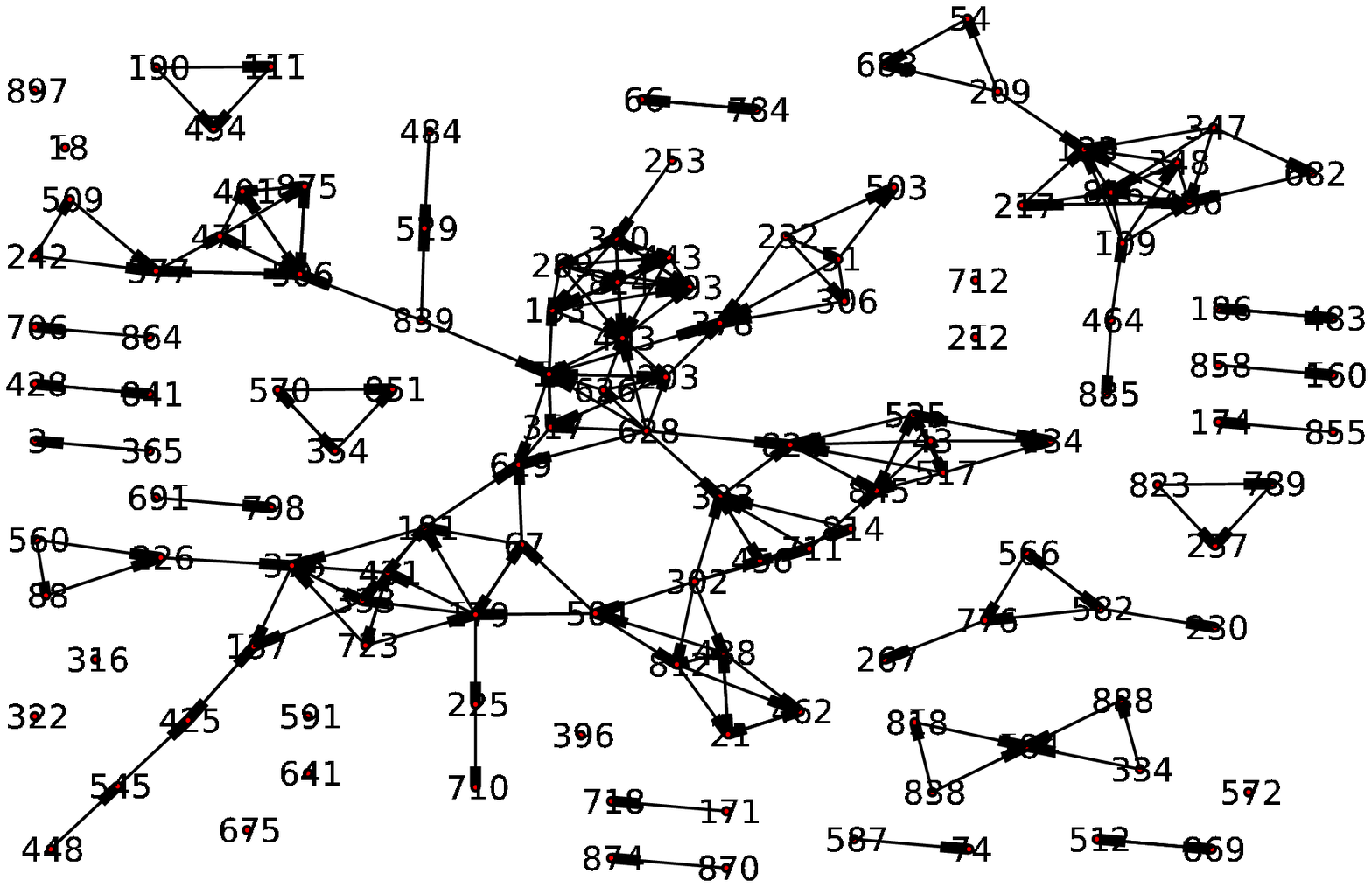}} \\ \\
  
    \resizebox{!}{0.11\textheight}{\includegraphics{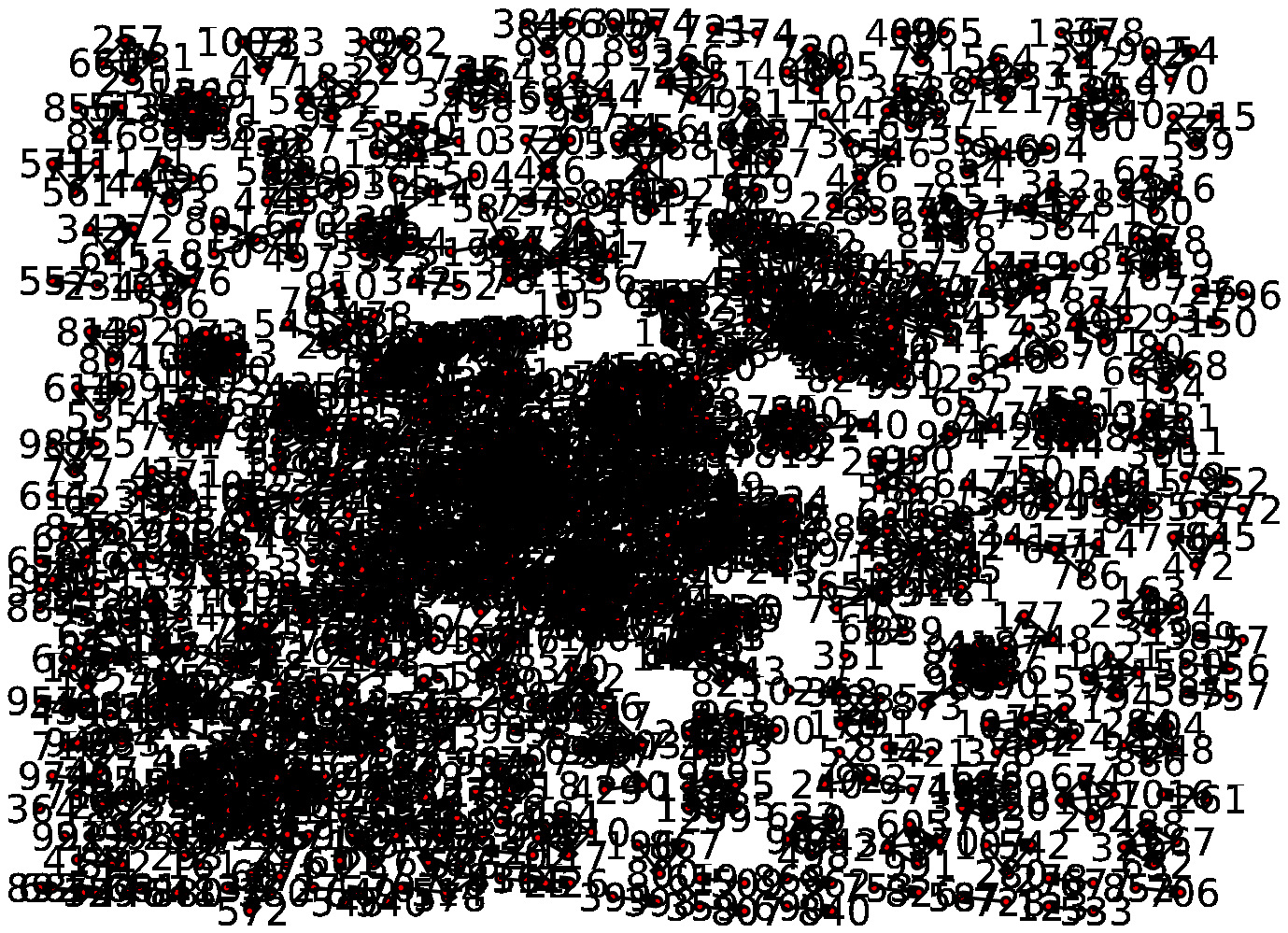}} & 
    \resizebox{!}{0.11\textheight}{\includegraphics{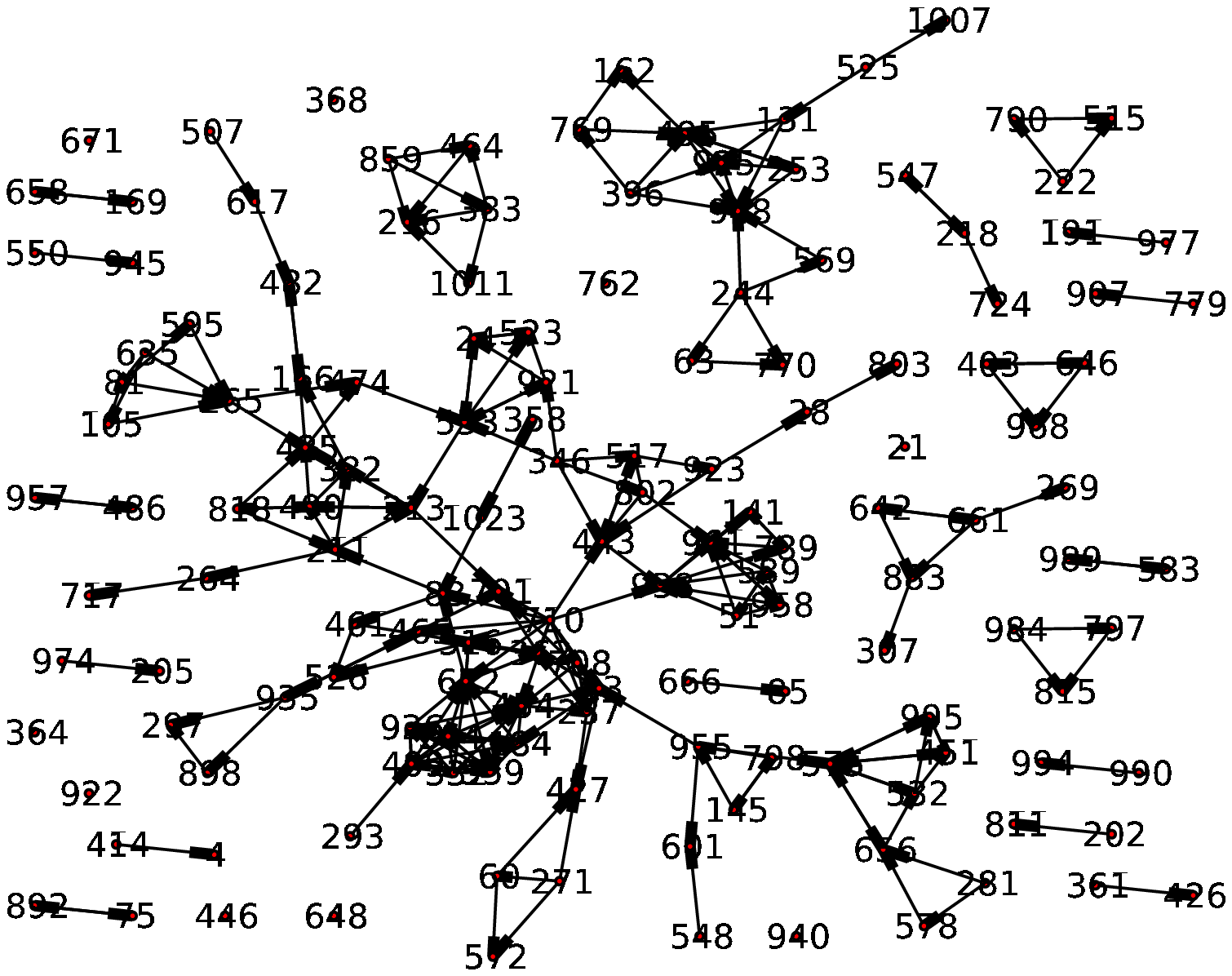}} \\ \\
  \end{tabular}
  \caption{Evolvement of coauthor network of ISWC from $2002$ to $2008$: each
    row shows the coauthor network and its backbone network; the left column
    shows the coauthor network, while the right column shows the backbone
    network.}
  \label{fig:coauthor-backbone-iswc-evolve}
\end{figure}

Fig. \ref{fig:coauthor-backbone-iswc-evolve} shows the evolution of coauthor
network of ISWC from $2002$ to $2008$. The coauthor networks are accumulated
year by year, that is, the coauthor network of year $n~(2002 \leq n \leq 2008)$
contains the coauthor relations from year $2002$ to year $n$.

The evolvement of coauthor network reflects the history of ISWC. More and more
researchers have taken part in the conference, while the nodes and links in
backbone networks are also changing. The following characteristics in the
evolvement of coauthor networks can be discovered:
\begin{enumerate}%
\item New researchers in the coauthor network often cooperate with the
  researchers that have published papers in ISWC conference, because the scales
  of modules in coauthor networks become larger year by year.

\item Scientific researchers are tending to cooperate with others. The evolution
  graph shows that the isolated nodes enter the connected components step by
  step.

\item Core researchers are tending to cooperate with each other. The number of
  researchers in the largest modules of backbone networks becoming larger and
  larger. This reflects the ``rich club'' phenomenon \cite{colizza2006drc} in
  scientific research.

\item Core researchers are active locally, and they have more cooperators than
  their neighbors. The roles of researchers in coauthor network are also
  changing: new researchers may become core researchers, while core researchers
  may become middle nodes or margin nodes.

\item The topological centers of the largest module are changing. The
  topological centers emerge through a voting-like mechanism. Table
  \ref{tab:iswc-centers} shows the topological centers.

\end{enumerate}

\begin{table}[!htb]
  \centering
  \caption{Topological centers of coauthor networks of ISWC from $2002$ to $2008$}
  \label{tab:iswc-centers}
  \begin{tabular}{|l|l|l|l|}
    \hline
    Year & \#Researcher & \#Cooperation & Topological Center
    \\ \hline \hline
    2002 & 99 & 174 & Katia P. Sycara \\ \hline 
    2003 & 262 & 510 & Katia P. Sycara \\ \hline 
    2004 & 393 & 877 & Steffen Staab\\ \hline 
    2005 & 570 & 1310 & Steffen Staab\\ \hline 
    2006 & 753 & 1872 & Guus Schreiber\\ \hline 
    2007 & 897 & 2290 & Guus Schreiber\\ \hline 
    2008 & 1024 & 2647 & Guus Schreiber\\ \hline 
  \end{tabular}
\end{table}

\section{Discussions}

\subsection{About the Topological Centrality}

The TC degree of a node reflects the geodesic distance to the nearest
topological center in the network. The value of TC degree has no definite
explanation, but it is different from the ranking results of PageRank. TC
degrees have close relation with the authority of nodes. Authoritative nodes
have higher TC degrees than its neighbors. The authority of a node reflects the
importance of a node in information propagation. The TC degrees are explainable
in communities. Core nodes have higher TC degrees than their neighbors. Isolated
resources have less influence in the global society.

Backbone networks can help study relations between resources of different types.
Backbone network of heterogeneous research networks connects important resources
in a research topic and important resources may be researchers, papers,
conferences, journals, institutions and publishers etc. This helps find and
recommend information. Furthermore, related information can be displayed by an
interactive visualization based browser.

In general complex networks, edges have no semantics. While in semantics-rich
networks, edges have semantic relations. Weights of nodes are affected by their
neighbors, and different relations have different effects. So it is necessary to
consider the influences of relations on the topological centrality calculation.
Relations can be assigned with different weights and participate the iterative
calculation as shown in Eq. (\ref{eq:semantic-tc}), where r is the relation of
link $e(i, j)$, $\omega_{r}$ is the weight of $r$ that affects the calculation
of TC in each iteration:

\begin{equation}%
\label{eq:semantic-tc}
  \left\{
    \begin{array}{ll}
      temp\_{\omega_{i}^{(t+1)}} & = \omega_{i}^{(t)} + \sum_{j=1}^{n}a_{ij} \omega_{r}\omega_{e(i,j)}^{(t))}\omega_{j}^{(t)} \\
      temp\_{\omega_{e(i,j)}^{(t+1)}} & = temp\_{\omega_{i}^{(t+1)}} + temp\_{\omega_{j}^{(t+1)}}
    \end{array}
  \right.
\end{equation}
Where $r$ is the relation of link $e(i, j)$, $\omega_{r}$ is the weight of $r$,
which affects the calculation of TC in each iteration.

An important characteristic is that the original topological centers may change
when we merge two networks into one by certain links and recalculate the
topological centers in the new network. For example, if we merge the coauthor
network with the citation network by the \textit{authorOf} semantic links, the
topological centers of the new network may not be simply the sum of the
topological centers in the coauthor network and those in the citation network.
Recalculation of topological centers can synthesize more relations, so this can
more accurately evaluate nodes. For example, authors can be evaluated by more
factors (e.g., number of publications, number of co-authors, number of
citations) in the new network than in the old networks. If applications require
to keep the old topological centers in the new network and avoid recalculation,
we can adopt the following strategy: find the relations (e.g.,
\textit{authorOf}) between the old topological centers and then compose the
corresponding old topological centers to form new topological centers. Such an
integrated topological centers can provide semantic relevant information
services (e.g., the authority author and his/her high impact papers can be
obtained at the same time) for applications in large network.

\subsection{Related Works}

General community discovery approaches are based on the connections between
vertices in a network. A fast community discovery algorithm in very large
network was proposed with approximate linear time complexity $O(nlog^2n)$, where
n is the number of nodes \cite{clauset2004fcs}. The general methods like GN
algorithm can be used to discover communities in weighted networks by mapping
them onto unweighted networks \cite{newman2004awn}.

Research and learning resources form a network, and the connections are the
relations among resources. Different from the communities in general complex
networks, semantic communities in the relational network were discovered
according to the roles of relations during reasoning on relations
\cite{zhuge08:_commun_and_emerg_seman_in}.

Many works are on the collaboration networks and citation networks of scientific
research. Most of them focus on the characteristics of collaboration networks.
For the structure of social science collaboration network, disciplinary cohesion
from $1963$ to $1999$ was studied \cite{moody2004sss}. The structure of
scientific collaboration networks including the shortest paths, weighted
networks, and centrality was studied \cite{bordons2000cns} \cite{newman2001ssc}
\cite{newman2001scn2}. Coauthor relations were used to study the collaborations
between researchers especially the mathematician, and the distribution of
relations between papers of Mathematical Review against the number of authors
was studied \cite{grossman1995pwk} \cite{melin1996src}. Relations between
researchers were analyzed in Ed{\"o}rs collaboration graph, and the shortest
path lengths between researchers were studied \cite{batagelj2000sae}.

Evolutions of the social networks of scientific collaborations in mathematics
and neuro-science were studied \cite{barabasi2002esn}. The research result shows
that the social network of collaboration network is scale-free; and, the node
separation decreases with the increase of connections.

Social network in academic research can be extracted from the webpages and paper
metadata provided by the online databases \cite{tang2007sne}; furthermore,
relations among researchers are mined in academic social networks
\cite{1402008}. Social structure in scientific research was studied based on the
citations \cite{white2004dcr}.

Citation relations between scientific papers, and the citation distribution of
papers was studied \cite{price1957nsp} \cite{seglen1992ss} \cite{redner1998pyp},
and shows that some papers are not cited at all, most papers are cited once,
while a little part of papers covers the references of most papers in a research
area.

Resources in research networks are ranked in \textit{Object} level. Research
resources were ranked by \textit{popRank} approach considering the mutual
influences between relevant resources \cite{1060828}. Object based ranking
approach can help search and recommend different resources such as papers,
conferences, journals and researchers etc.

Researchers and papers are often ranked in coauthor network and citation network
respectively. A co-ranking framework of researchers and papers was proposed, in
which researchers and papers were ranked in a heterogeneous network combining
the coauthor network and citation network by coauthor relations
\cite{zhou2007cra}.

Our approach is different from the existing approaches in the following aspects:

\begin{enumerate}%
\item We distinguish the roles of nodes by topological centrality, and then discover
  the communities by roles of nodes. Global communities and local communities are
  discovered based on the roles of nodes. So our approach is based on role rather
  than only on the connections. Although the topological centrality degrees of
  nodes and edges are calculated considering connections between nodes, the
  topological centrality degrees of neighbor nodes have influences on each other
  at the same time. The role based community discovery approach is fit for the
  research networks, and can discover communities in tree-like networks that are
  hard to discover by general community discovery approaches such as GN
  algorithm.

\item We have built the backbone networks for coauthor networks and citation
  networks, and the evolution characteristics of backbone networks have been
  studied. PageRank algorithm can also find the local core nodes, but it has no
  way to connect most of the core nodes into a backbone network, because it is
  hard to choose the connecting nodes between the core nodes by the PageRank
  values. While topological centrality degrees of nodes can choose the core
  nodes and connect them into a connected backbone network as more as possible,
  because the core nodes include the community central nodes and important nodes
  connecting different communities. The backbone network construction approach
  is also based on the topological centrality. The approach can be applied not
  only in the research networks with single resource type but also those with
  multiple resource types.

\end{enumerate}

\section{Conclusion}

This paper first proposes the notion of topological centrality and the
calculation approach to reflect the topological positions of nodes and edges in
a network, and then studies its applications in discovering communities and
building the backbone network in scientific research networks. Research
communities can be discovered according to the roles of nodes distinguished by
topological centrality degrees. We also propose an approach to building the
backbone network by using the topological centrality. Experiments on real
research network and simulation networks show the feasibility and effective of
our approaches.

\ifCLASSOPTIONcompsoc

\section*{Acknowledgments}
\else

\section*{Acknowledgment}
\fi

This research work was supported by the National Basic Research Program of China
(Project No. 2003CB317000).

\ifCLASSOPTIONcaptionsoff \newpage \fi

\end{document}